\documentclass[aps,prb,preprint,superscriptaddress]{revtex4-1}

\usepackage[english]{babel}
\usepackage[version=4]{mhchem}
\usepackage{graphicx}
\usepackage{xcolor}

\newcommand{\cblu}[2][black]{\textcolor{#1}{#2}}

\def \pst {Pb$_{1-x}$Sn$_x$Te}

\newcommand{\angstrom}{\mbox{\normalfont\AA}}

\begin{document}


\title{Tunable chiral anomaly in electron magnetotransport in the Weyl semimetallic \ce{Pb_{1-x}Sn_xTe}:Cr alloy }


\author{A. Królicka }
\affiliation{Institute of Physics,Polish Academy of Sciences, al. Lotników 32/46, 02668 Warsaw, Poland}

\author{E. Łusakowska }
\affiliation{Institute of Physics,Polish Academy of Sciences, al. Lotników 32/46, 02668 Warsaw, Poland}

\author{M. Matusiak  }
\affiliation{Institute of Low Temperature and Structure Research, Polish Academy of Sciences, Okólna 2, 50422 Wrocław, Poland}

\author{A. Mirowska  }
\affiliation{ENSEMBLE3 Centre of Excellence for Nanophotonics, Advanced Materials and Novel Crystal Growth-Based Technologies, Wólczyńska 133, 01919 Warsaw, Poland}

\author{A. Łusakowski  }
\affiliation{Institute of Physics,Polish Academy of Sciences, al. Lotników 32/46, 02668 Warsaw, Poland}

\author{T. Story }
\affiliation{Institute of Physics,Polish Academy of Sciences, al. Lotników 32/46, 02668 Warsaw, Poland}
\affiliation{International Research Centre MagTop, Institute of Physics. 
	Polish Academy of Sciences, al. Lotników 32/46, 02668 Warsaw, Poland
}

\author{K. Dybko }
\affiliation{Institute of Physics,Polish Academy of Sciences, al. Lotników 32/46, 02668 Warsaw, Poland}
\affiliation{International Research Centre MagTop, Institute of Physics. 
	Polish Academy of Sciences, al. Lotników 32/46, 02668 Warsaw, Poland
}

\date{\today}

\begin{abstract}

We study magnetotransport properties of semiconductor substitutional alloy \ce{Pb_{1-x}Sn_xTe}, known to exhibit Sn-content dependent properties of topological crystalline insulators with a semimetallic zero-gap state at a specific band inversion point. We experimentally verify the theoretically predicted role of chemical disorder in this multivalley electron system, which leads to sequential band inversions in various valleys and places the Fermi level close to the pairs of Weyl nodes, as identified in the density functional theory calculations. Doping with mixed-valence Cr resonant impurities enables exploitation of the unique properties of dopant resonant states, which provide an effective means of tuning carrier concentration. The combination of these two effects results in the pinning of the Fermi level in the vicinity of the nodal touching points across a wide range of composition. To address the above issues, we grow Bridgman bulk crystals of \ce{Pb_{1-x}Sn_xTe} heavily doped with chromium and covering the full range of tin (0 $\leq x \leq$ 1), i.e. spanning both the topological crystalline insulator and trivial electronic regimes. We observe the emergence of the three dimensional (3D) Weyl semimetal phase over a range of Sn compositions, namely for $0.25 < x < 0.45$. We provide magnetotransport evidence for this and verify the relationship between the magnitude of the experimentally determined Berry curvature and the electrical properties of these materials. Quantum transport regime observed in magnetoresistance is also independently confirmed by thermal conductivity measurements. For one of our samples we observe the striking phenomenon of the room-temperature re-entrance of the chiral anomaly, in addition to its occurrence in the 93–150~K temperature range. We also note a pronounced chiral anomaly-induced negative magnetoresistance, reaching up to 75$\%$. Consequently, the proposed approach creates favorable conditions for realizing the Weyl semimetal phase not just for one material with one specific composition, but in tunable complex IV-VI compounds.

%
	  

\end{abstract}


\maketitle

{\bf \large Introduction}\\
The discovery of topological phases of matter has fundamentally transformed modern condensed matter physics. Weyl semimetals (WSMs), which belong to this rapidly developing field, differ substantially from both topological insulators (TIs) \cite{bernevig_quantum_2006,fu_topological_2007,zhang_topological_2009,hasan_colloquium_2010,moore_birth_2010,ando_topological_2013} and topological crystalline insulators (TCIs) \cite{dziawa_topological_2012,tanaka_experimental_2012,xu_observation_2012,ando_topological_2013,ando_topological_2015}. Topological insulators exhibit an insulating bulk and have metallic surface states protected by certain symmetry: topological insulators - time-reversal symmetry \cite{fu_2011_TI_TRS}, and topological crystalline insulators - mirror crystalline symmetry \cite{fu_2011_TI_TRS}. On the other hand, Weyl semimetals remain gapless in the bulk and their surface states emerge when one of the symmetries is broken, either time-reversal or space inversion. This breakdown occurs under appropriate external perturbations including - structural distortion, lattice disorder, pressure, impurity doping, magnetic field or temperature. As a consequence of this symmetry breaking, these materials may evolve into Weyl semimetals through band inversion processes. This occurs via the topological phase transition between a TI/TCI and a normal insulator, as Weyl semimetals always occupy the intermediate position between these two states. Systems based on \ce{Pb_{1-x}Sn_xTe} or SnTe provide well-known examples of such phases \cite{lusakowski_alloy_2018,wang_digging_2019,liang_pressure-induced_2017}. For instance Liang et al. demonstrate that applying pressure to \ce{Pb_{1-x}Sn_xTe}:In, material undergoes a pressure-driven transition from TCI to a topological metallic phase associated with Weyl nodes. This results in splitting one degenerate Dirac cone where opposite-chirality fermions overlap into two nondegenerate Weyl cones of opposite chiralities separated in momentum space. This constitutes another characteristic of WSMs which distinguishes them from TIs/TCIs. Both systems exhibit a topological nature, expressed through their spin texture, however in each case it appears differently. Finally, two/three dimensional (2D/3D) cones (2D WSMs/3D WSMs) emerge connected by the so called Fermi arcs - topological states of the WSMs located on surface but being connected by bulk.

Introducing chemical disorder and other effects, that break local crystal symmetries and split valley degeneracy at high symmetry points of the Brillouin zone (BZ) may lead to significant changes, e.g. in the band inversion process in comparison with TIs/TCIs. According to the classical convention, these materials are usually considered within the virtual crystal approximation (VCA), where, for a given composition, the transition from trivial to topological phase occurs. In the case of \ce{Pb_{1-x}Sn_xTe} it is approximately 35 at.\% of Sn at 4.2 K. Then, the band structure shows inverted order of the $L_6^+$ and $L_6^-$ bands and the energy gap in the four $L$ points of the BZ becomes closed. Recent reports point to a different approach to band inversion processes taking place in 3D WSM alloys. Of particular importance are two papers: Wang et al. \cite{wang_digging_2019}  demonstrated that chemical disorder on cation position in \ce{Pb_{1-x}Sn_xSe} alloys can induce an intermediate WSM phase during the sequential band inversion in four "L" valleys. Similar conclusion was reached earlier by Lusakowski et al. \cite{lusakowski_alloy_2018} in the study on \ce{Pb_{1-x}Sn_xTe} alloys. In particular, the work shows how alloy disorder broadens the transition region between trivial and topological phases, producing an extended transition regime associated with nontrivial topology. The WSM phase predicted in this work is located in the region of about $\pm$10 at.\% of Sn concentration, starting from the point of the band gap closing in the VCA ($x$=0.35). From physical perspective, both studies describe essentially the same mechanism - disorder transforms a sharp and concurrent (VCA-like) band inversion into a sequential one with the finite composition range where WSM-like states may emerge.  

Besides the aforementioned exotic topological surface states, quasiparticles analogous to massless chiral Weyl fermions are realized in these materials - species originally introduced by Weyl back in 1929 during his work on the relativistic quantum mechanics and quantum field theory  \cite{weyl_elektron_1929}. The main highlight of this work in our context is that since a bulk energy-gap, responsible for the mass-term of the Dirac equation is closed in WSMs this may be converted into a massless Weyl equation which results in linear energy dispersion in the vicinity of the two Weyl nodes \cite{zeljkovic_dirac_2015,orbanic_quantum_2017}. In contrast to non-relativistic electrons present in ordinary metals and semiconductors with mainly parabolic or Kane-type energy dispersion, electrons in WSMs have spin locked to the momentum, and thus do not undergo back-scattering by spin-independent processes, in which they resemble TIs/TCIs. As a result, they  behave like massles relativistic Dirac fermions, giving rise to large non-saturated magnetoresistance, and hence very high mobilities (energy-independent relaxation time) in these materials.

Several decades after Weyl's achievements, Adler, Bell and Jackiw demonstrated that due to the nontrivial nature of Weyl fermions, the chiral symmetry is no longer preserved, leading to the Adler-Bell-Jackiv anomaly. Nielsen and Ninomiya demonstrated that lattice systems necessarily contain Weyl fermions in pairs of opposite chirality and argued that condensed matter systems could exhibit signatures of the chiral symmetry breakdown under parallel electric and magnetic fields. This work laid the foundation for understanding anomalous magnetotransport effects in Weyl semimetals, showing that the chiral anomaly observed in condensed matter physics is the analogue of the Adler-Bell-Jackiv anomaly in high-energy physics. Chiral anomaly manifests itself as a non-conservation of chiral charge between two Weyl nodes of opposite chiralities when applying parallel electric and magnetic fields ($E \parallel B$) to the material. It is reflected experimentally as negative magnetoresistance ($MR_\parallel$) and observation of chiral anomaly is treated as one of the crucial signatures confirming presence of the 3D Weyl semimetal phase. An interesting contribution to this topic was provided especially in several publications \cite{son_chiral_2013,kim_BiSb_2013,burkov_chiral_2014,hirschberger_chiral_2016}. Chiral anomaly-induced negative $MR_\parallel$ has been observed for the first time in \ce{Bi_{1-x}Sb_x}, $x=0.03$ in 2013 \cite{kim_BiSb_2013}. Soon after, researchers also reported its presence for Dirac and Weyl semimetals such as: \ce{Cd_3As_2} \cite{li_Cd3As2_giant,zhang_room-temperature_Cd2As3_2017} , \ce{NaBi_3} \cite{xiong_na3bi_2015} , TaAs \cite{xiong_TaAs_2015,huang_observation_2015,zhang_signatures_2016}, TaP \cite{hu__2016_TaP}, $\alpha$-Sn \cite{polaczynski_3d_2024} . 

Interestingly, both mentioned mechanisms - positive linear, non-saturating  $MR_\perp$ and negative $MR_\parallel$, may be the consequence of one and the same phenomenon - both  may originate from the system entering the extreme quantum limit, i.e. where all electrons occupy only the lowest Landau level. These findings were made independently by two research groups - the model of Argyres and Adams \cite{argyres_ultraquantumlimit_1956,zhang_signatures_2016}  addresses longitudinal magnetoresistance and may lead to negative $MR_\parallel$, whereas second model of quantum magnetoresistance, implemented by Abrikosov \cite{abrikosov_quantum_1998,wang_linearMR_2012}, predicts a non-saturating linear  $MR_\perp$.

A major conceptual breakthrough occurred when Weyl nodes were recognized as topologically protected monopole and antimonopole of Berry flux in momentum space.  One of the earliest theoretical predictions of the Weyl semimetal phase in solids was provided for \ce{Y_2Ir_2O_7} in 2011. Its experimental confirmation followed shortly afterwards with the observation of Weyl fermions and Fermi arcs using the Angular-Resolved Photoemission Spectroscopy technique in TaAs compounds. The Berry curvature associated with Weyl nodes plays a central role in determining the physical properties of these materials, strongly influencing both: longitudinal charge transport, reflected in the chiral anomaly, and Hall responses. Especially important development was the prediction of the nonlinear quantum Hall effect in 2015 \cite{sodemann_quantum_2015}. The nonlinear quantum Hall effect is a second-order response to an AC driving current. This effect is realized  when the external magnetic field is absent  and the output signal exhibits a quadratic dependence on the input signal AC driving current. The resulting asymmetric Berry curvature distribution originates from the broken inversion symmetry under which electrons at momenta k and –k possess opposite Berry curvature. Hence, charge states for k and -k will be occupied differently leading to the net Berry curvature. This discovery significantly broadens the scope of topological transport phenomena and demonstrates that nonlinear electronic responses may emerge purely from band geometry.

Another important area of investigation of WSMs is anomalous Hall effect (AHE). Although the effect has been first observed yet in the nineteenth century \cite{Hall1879OnAN}  and was for long associated solely with ferromagnetism, the modern understanding of AHE as the effect originating from the Berry curvature and momentum space magnetic monopoles developed only after the 2000s \cite{Fang,Haldane,sinitsyn_disorder_2005,nagaosa2006,Wang,werpa_2011}. The review article "Anomalous Hall effect" \cite{Nagaosa2010} published in 2010 synthesizes previous findings indicating that the anomalous contribution to the Hall conductivity has an intrinsic nature and can be described through the Berry curvature integral over the occupied electronic bands throughout the Brillouin zone. A few years later \cite{liang_pressure-induced_2017} the scientific group from Princeton observed a nonlinear Hall contribution ($\sigma_{yx}$) beyond the ordinary linear Hall effect in \ce{Pb_{1-x}Sn_xTe}, which they assigned to the large Berry curvature situated near the Weyl nodes. In their paper they extract this anomalous contribution and estimate the value of the Berry curvature.

Our recent studies were devoted to the band inversion and band-offsets in Cr-doped \ce{Pb_{1-x}Sn_xTe} bulk crystals covering whole range of Sn composition $(0 \leq x \leq 1)$ \cite{krolicka_cr_2025}. In that study we also determine the position of the Cr resonant level over the entire \ce{Pb_{1-x}Sn_xTe}:Cr phase diagram $(0 \leq x \leq 1)$. This is possible due to the Fermi level pinning effect, whose application has so far been limited solely to improving thermoelectric performance \cite{heremans_enhancement_2008,heremans_resonant_2012,krolicka_comparison_2017,krolicka_synthesis_2017}. Since \ce{Pb_{1-x}Sn_xTe} is in its trivial or TCI phase at low and high Sn contents, respectively, we were able to analyze its transport properties using a classical approach. Nevertheless, for samples with Sn contents in the intermediate composition range: $0.25~<~x~<~0.45$, we observe highly unusual transport properties - sudden drop of carrier concentration and Fermi energy. This behavior may be understood in terms of compositional-disorder driven transition to WSM phase according to earlier theoretical predictions \cite{lusakowski_alloy_2018,wang_digging_2019,lusakowski_band_2021}, instead of the VCA approach with band inversion point situated at $x=0.35$  in \ce{Pb_{1-x}Sn_xTe}. Subsequent high field magneto-transport measurements of the resistivity tensor confirm that our system realizes the Berry curvature induced Weyl semimetal phase. Especially, this is evidenced by the observations of the intrinsic anomalous Hall effect in perpendicular magnetic field and chiral anomaly in parallel electric and magnetic fields. Also, the studies of the quantum nonlinear Hall effect and the detection of the second harmonic signal do confirm the existence of the Weyl semimetal phase in the investigated system. Additionally, quantum transport and especially the Shubnikov-de-Haas oscillations, detected in several of our samples, due to the good crystal quality, provided crucial evidence for the $\pi$ Berry phase in the studied samples.

{\bf \large Results and discussion \label{Results}}

{\bf The role of alloy disorder and resonant dopants}\\
To date a great disadvantage of many projects on (3D) Weyl semimetals has been the lack of precise control of the carrier concentration. Hence, previous studies focused mostly on materials whose compositions lie within the band-crossing region and are 3D WSMs in the whole temperature range, like Pb$_{0.48}$Sn$_{0.52}$Te thin films \cite{nishijima_ferroic_2023,zhang_giant_2022}  or (Pb$_{0.5}$Sn$_{0.5}$)$_{0.94}$In$_{0.06}$Te single crystals \cite{liang_pressure-induced_2017}.

\begin{figure} [h]
	\includegraphics[width=1\textwidth]{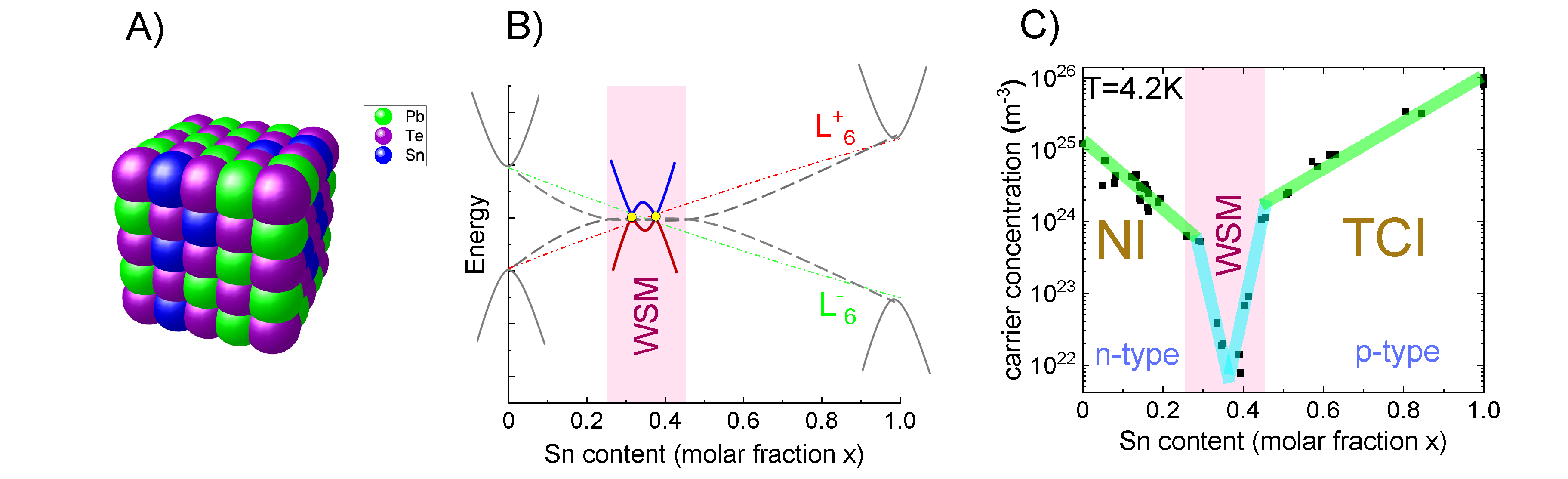}%
	\caption{\label{fig:F1} (A) Schematic representation of the NaCl unit cell with Sn-induced substitutional alloy disorder (Sn atoms substitute Pb atoms on cation sites). (B) Band structure of \ce{Pb_{1-x}Sn_xTe} over the full composition range (0 $\leq x \leq$ 1). The theoretically predicted region (shaded in pink) with the sequential band inversion process leading to extended range of the Weyl states (blue and red bands with yellow dots) with closed gap, is schematically highlighted (grey dashed line). In contrast, the simplified virtual crystal approximation (VCA) model predicts a topological phase transition only at a single composition. The $L_6^+$ and $L_6^-$ bands are highlighted in red and green, respectively. The shapes of the bands are symbolic. (C) Carrier concentration as a function of Sn content $x$ for our \ce{Pb_{1-x}Sn_xTe}:Cr samples (0 $\leq x \leq$ 1) at 4.2 K. A pronounced minimum in carrier density is observed in the composition range corresponding to the Weyl semimetal (WSM) phase. The other two regions correspond to the normal insulator (NI) and the topological crystalline insulator (TCI) phases, for samples with low and high Sn contents, respectively.Previously published by the author; see Ref. \cite{krolicka_cr_2025} for linear-scale representation and band inversion analysis.} 

\end{figure}

Within the framework of this work we exploit two unique features of \ce{Pb_{1-x}Sn_xTe} to overcome these difficulties. First,  following the theoretical proposal by Lusakowski et al. \cite{lusakowski_alloy_2018}, we exploit alloy disorder in Pb$_{1-x}$Sn$_{x}$Te crystals. This is schematically shown in Fig. \ref{fig:F1}A by illustrating substitution of Pb ions by Sn on cation positions. The grown crystals cover full range of Sn composition (0 $\leq x \leq$ 1), thereby extending the band-inversion regime, i.e. the region where the band-gap is closed, from single point to finite composition window. This allows broadening of the range of Sn composition for which the 3D WSM phase is observed (Fig. \ref{fig:F1}B). The second and key step is the incorporation of a resonant dopant - Cr into our crystals. As it is well established \cite{kaidanov_resonant_1992}, a very interesting property of resonant dopants with mixed valence states is the ability to pin the Fermi level ($E_F$) position at specific energy in the band structure, simultaneously leading to an increase in electronic homogeneity. When it comes to Cr, it is of particular interest, since as we have demonstrated thoroughly in our previous works \cite{story_transport_1992,lusakowski_bi_2025,krolicka_cr_2025}, mixed valence ions of chromium (Cr$^{2+/3+}$) form resonant donor centers in the conduction band ($CB$) of PbTe but located in the valence band ($VB$) in SnTe, thereby pinning the Fermi level in both these materials. As a result, the combination of these two steps causes the Cr$^{2+/3+}$ level to be located close to the vanishing band gap at the band inversion point and the WSM region spans the intermediate range of compositions.

The carrier concentration of our samples extracted from low-field Hall data is summarized in Fig. \ref{fig:F1}C. As can be seen, in the vicinity of the gap-closing point ($\pm$ 10 at.\% of Sn content), a pronounced change in the trend of carrier concentration is observed. A strong suppression of the carrier concentration takes place, yelding the record low value $n=1$ x $10^{16}$~cm$^{-3}$ for $x = 0.38$ (sample I), which suggests/implies that $E_F$ resides in the close proximity of the  band touching points, confirming that our system is in the quantum regime, i.e. placed close to the Weyl nodal points \cite{zhang_signatures_2016}. This unique trend  extends over approximately 20 at.$\% $  of Sn composition ($0.25<x<0.45$), which is consistent with the work of Lusakowski et al. \cite{lusakowski_alloy_2018}. Due to chemical disorder (lower local symmetry), the WSM state with a zero gap at four non-equivalent $L$ points emerges over a range of Sn compositions.

According to Fig.  \ref{fig:F1} and to the above analysis, our system may be devided into three particular phases: (I) normal insulator (NI) ($x<0.25$, $E_F$ pinned in the $CB$), (II) Weyl semimetal (WSM) ($0.25<x<0.45$, $E_F$ pinned in the close proximity of the Weyl nodal points) and (III) topological crystalline insulator (TCI) ($x>0.45$, $E_F$ pinned in the $VB$). The boundaries between these regimes define composition-induced topological phase transitions. Therefore, as stated at the beginning of this chapter, we demonstrate that by systematically lowering carrier concentration of pure \ce{Pb_{1-x}Sn_xTe} (due to appropriate amount of Cr dopant) we are able to tune the Fermi level and the emergence/disappearance of a 3D WSM phase, which we further substantiate using different mechanisms such as the chiral anomaly and the quantum nonlinear Hall effect.\\

{\bf Transport measurements}

\begin{figure} [h]
	\includegraphics[width=1\textwidth]{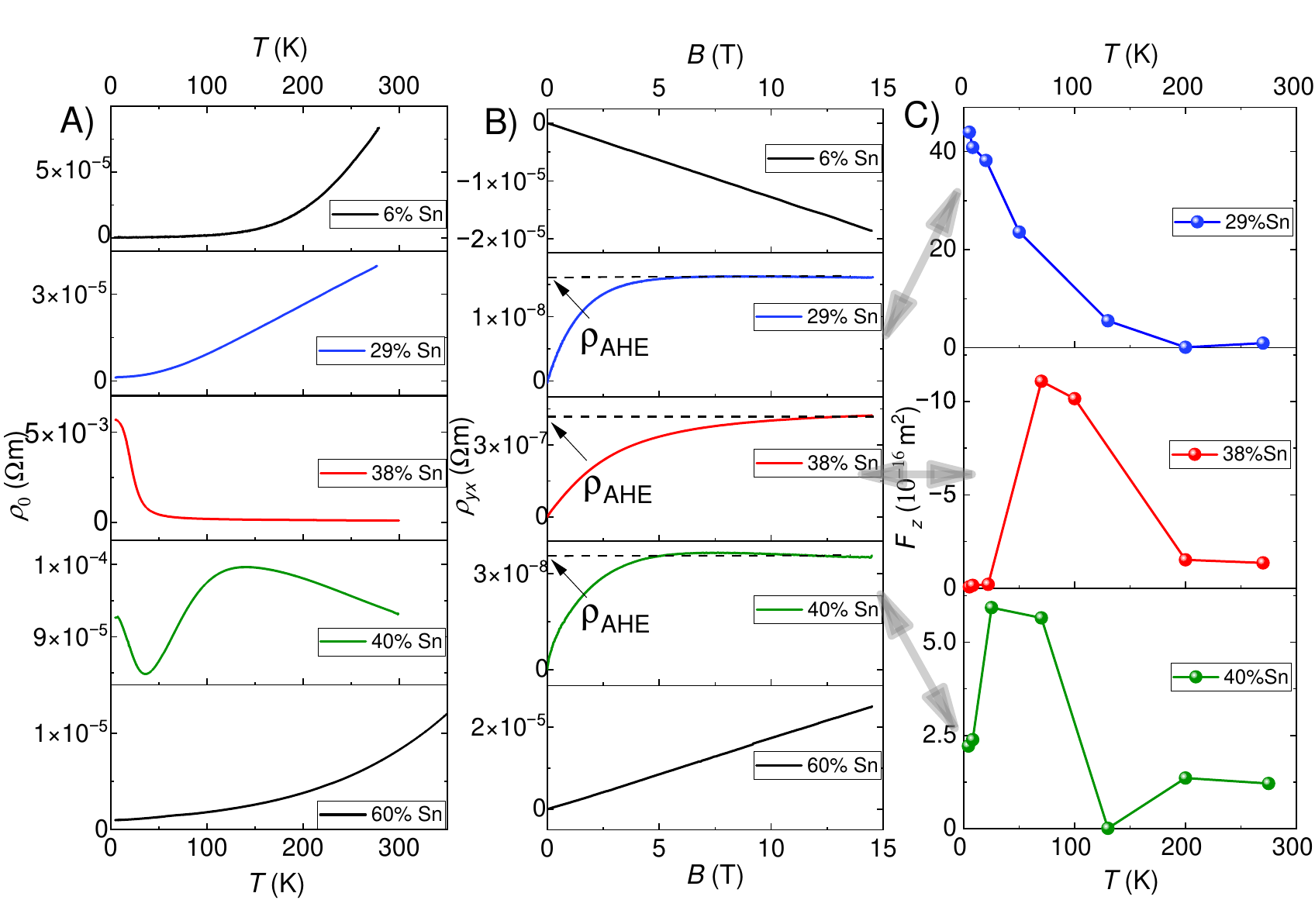}%
	\caption{\label{fig:F2} (A) Zero-field resistivity $(\rho_0)$ versus temperature for several \ce{Pb_{1-x}Sn_xTe}:Cr samples (0.06 $\leq x \leq$ 0.6), covering three regions of Sn composition: normal insulator (NI), Weyl semimetallic (WSM), topological crystalline insulator (TCI). (B) The corresponding Hall resistivity $(\rho_{yx})$ versus magnetic field dependencies for chosen representative temperatures ($T=50$ K for $x=0.29$, $T=100$ K for $x=0.38$; for other samples, results obtained under liquid helium conditions are presented). Results reveal linear $\rho_{yx}$ versus magnetic field dependencies for samples from regions I and III and anomalous behavior (IAHE) for three samples from the WSM region ($x=0.29$, $x=0.38$, $x=0.4$). (C)  $\mathcal{F}$$_z$=$\sigma_{AHE}$/$n$ versus temperature determined for three samples from the WSM region.  $\mathcal{F}$$_z$ is proportional to the Berry curvature. }
	
\end{figure}

{\em Intrinsic anomalous Hall effect and the Berry curvature}\\
Apart from the composition dependent topological/trivial phase transition, we also observe a temperature-driven topological/trivial phase transition, due to temperature-induced changes in the band gap (and lattice constant). Figure \ref{fig:F2} shows our transport measurement results for several representative samples characterizing three distinct phase regions, shown in Fig.~\ref{fig:F1}. Indeed, for samples with tin compositions \mbox{$0.25<x<0.45$} we observe deviations in the temperature-dependent zero-field resistivity $(\rho_0)$   (Fig.~\ref{fig:F2}A) as compared to ordinary semiconductors and TCIs. While all our samples outside the WSM composition range exhibit behavior of highly degenerate semiconductors with metallic-like conductivity, samples in the WSM regime exhibit also semiconducting or temperature independent behavior.

Together with the corresponding Hall resistivity $\rho_{yx}$ (Fig. \ref{fig:F2}B), the performed $\rho_0(T)$ measurements provide additional way to distinguish samples within and outside the WSM regime. This is because of Hall response of the WSMs, which contains the nonlinear contribution, resulting from the Berry phase acquired by electrons evolving in the k-space. Confirmation that this contribution is of intrinsic nature comes from the fact, that in contrast to the traditional AHE observed in ferromagnets, Hall signal in our samples vanishes at zero field, indicating the absence of the hysteresis loop characteristic of ferromagnets. In other words, in the vicinity of the zero magnetic field, Hall response is determined solely by the contribution of charge carriers, residing in topologically trivial states ($\rho_{N}$). This, so called, intrinsic anomalous Hall effect (IAHE) provides a direct way to determine the Berry curvature of the material's band structure. To this end, we analyze the intrinsic contribution to the anomalous Hall effect following well established approaches \cite{liang_pressure-induced_2017}. First, we extract the anomalous Hall resistivity $\rho_{AHE}$ from the linear extrapolation of the high-field Hall resistivity curve, as presented in Fig. \ref{fig:F2}B. The Berry curvature scales with the ratio of the anomalous Hall conductivity $\sigma_{AHE}$ to the carrier concentration. Our results are plotted in Fig. \ref{fig:F2}C and for a more detailed analysis see Supplementary Note S.I. As we mentioned, the carrier concentration (plotted in Fig. \ref{fig:F1}) can be calculated from the linear slope of the low-field Hall resistivity, using the Drude model. Since at low magnetic fields (where the $\mathcal{F}$$_z$ is negligible) our $\rho_{yx}$ depends linearly on $B$, therefore it is safe to use single carrier density model \cite{liang_pressure-induced_2017} (see Supplementary Note S.II for more detailed explanation). Our analysis reveals that samples with compositions from the WSM region exhibit enhanced $\mathcal{F}$$_z$ (Fig.\ref{fig:F2}C), while for samples with the trivial compositions $\mathcal{F}$$_z$ = 0. This behavior is consistent with the presence/absence of the Weyl nodes, which act as sources and sinks of the Berry flux. This is indicated by the decrease of the magnitude of this effect as the system is tuned away from the WSM regime, either by changing the temperature or composition.\\ 

{\em Chiral anomaly and its relation with other transport results}\\ 
\begin{figure} [hp]
	\vspace*{-1cm}
	\includegraphics[width=1\textwidth]{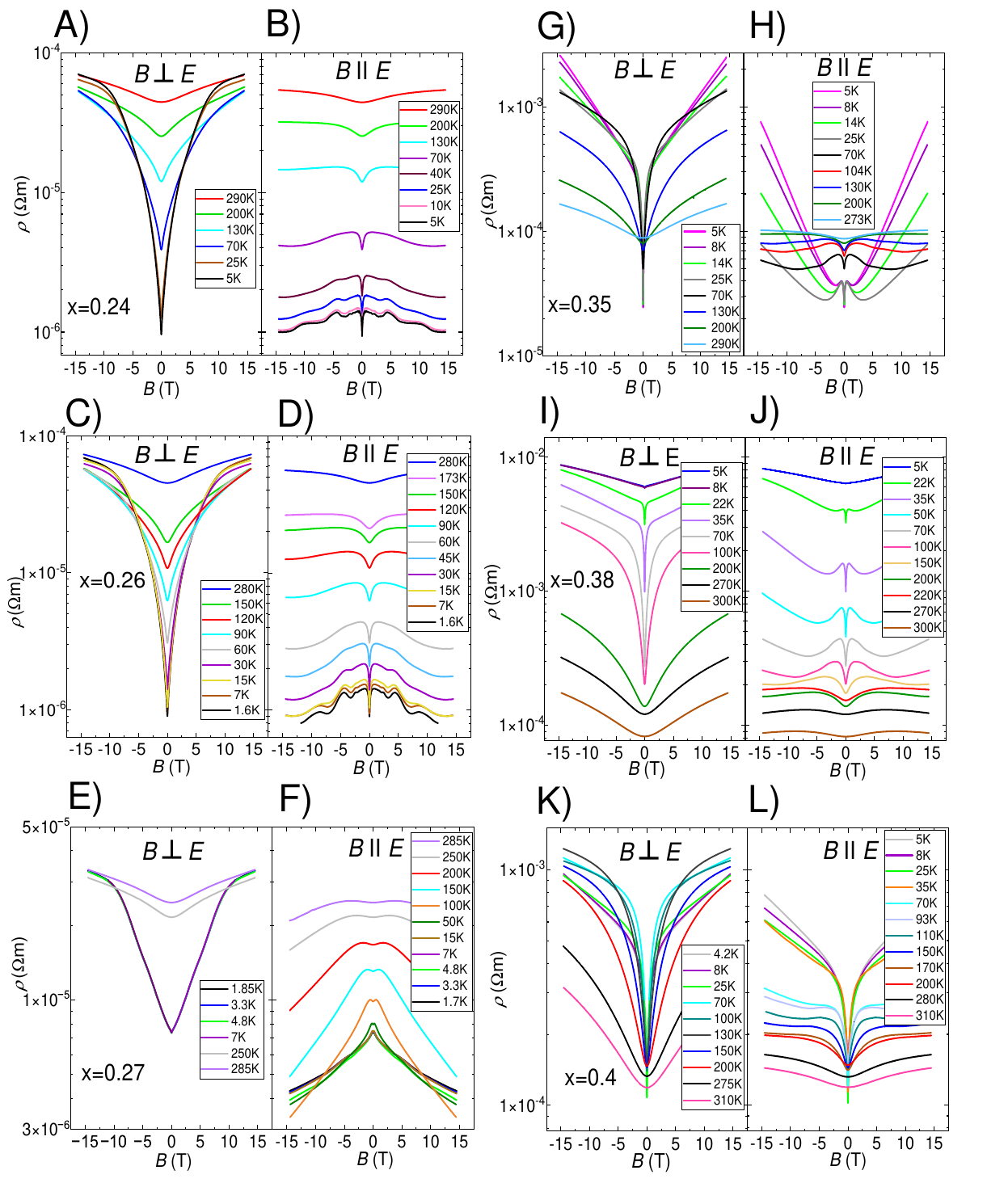}%
	\caption{\label{fig:F3} Resistivity as a function of magnetic field for representative samples from the WSM composition region (\ce{Pb_{1-x}Sn_xTe}, $0.25<x<0.45$) measured at various temperatures. Measurements were conducted in two configurations: with the current direction perpendicular to the magnetic field $\bf{B \perp E}$ (presented in panels: a, c, e, g, i, k) and with the current direction parallel to the magnetic field $\bf{B \parallel E}$ (presented in panels: b, d, f, h, j, l).   }
\end{figure}
For all \ce{Pb_{1-x}Sn_xTe}:Cr samples (0 $\leq x \leq$ 1) magneto-transport measurements were performed over a wide temperature range, for various magnetic field orientations with respect to the applied electric field. We do not include all these results here. Within this work we focus on measurements for samples from the WSM region $(0.25<x<0.45)$. Figure~\ref{fig:F3} presents magnetic field dependent resistivities of the WSM samples measured at two orientations - with magnetic field direction perpendicular ($\rho_\perp$) or parallel ($\rho_\parallel$) to the applied electric field. When $\bf{B \perp  E}$ samples reveal large unsaturated positive magnetoresistance $MR_\perp$. The results of the maximum $MR_\perp$ and  $MR_\parallel$ for all samples from Fig. \ref{fig:F3} are presented in Supplementary Note~S.III. Although our results deviate from maximum values currently reported in the literature \cite{okazaki_extremely_MR_2026}, they point to the linear dispersion relation of the topological charge carriers. In contrast, when $\bf{B \parallel E}$, a pronounced negative contribution to magnetoresistance response is observed, which is a hallmark of a chiral anomaly in WSM systems. This negative contribution is often preceded by the positive contribution, broadening with temperature, which may be attributed to the weak antilocalization (WAL) \cite{checkelsky_quantum_2009,Chen,kim_thickness-dependent_2011,shen_revealing_2015}, as it has been already documented in the literature on WSMs \cite{li_negative_2017,liang_experimental_tests2018,liang_pressure-induced_2017}. WAL occurs  in the systems with large spin-orbit coupling or in topological systems with Berry phase $\pi$, i.e. when the destructive interference of the electron paths takes place resulting in forbidden backscattering processes. Besides, when $\bf{B \perp  E}$, the resistivity versus $B$ dependencies exhibit "WAL" behavior in the low-field range and it is always followed by the positive MR. It points to the Berry phase since, if there was no $\pi$ Berry phase, WAL-like behavior would be followed by the weak localization (WL), regardless of the measurement configuration. The "WAL scenario" is also supported by fitting the appropriate formula for the 3D WSM system magnetoresistance, which is proportional to $B^2$ and to $\sqrt{B}$, at lowest and higher magnetic fields, respectively \cite{lu_weak_2015,zhang_signatures_2016}. For the in-depth analysis see Supplementary Note S.IV. The observed negative magnetoresistance contribution decreases and eventually disappears as the system undergoes crossover to another phase (NI or TCI). This can be easily seen for each sample from Fig. \ref{fig:F3}, for instance in Fig. \ref{fig:F3}D, where we observe chiral anomaly in the 1.6-120 K temperature range, hence the system is in the WSM phase. For higher temperatures the negative contribution vanishes indicating the transition to the NI phase. The temperature of the transition differs from sample to sample and depends on the resulting   Fermi level position. For the majority of the investigated samples, the highest chiral anomaly contribution is observed in the lowest temperature range and it decreases with increasing temperature. The strong magnetic field dependence of the chiral anomaly for several samples has already been reported in studies on Weyl semimetals \cite{li_negative_2017, hu__2016_TaP}. According to Ref. \cite{xiong_na3bi_2015}, such behavior may be attributed to the extreme sensitivity of the negative magnetoresistance to even slight misalignment between magnetic and electric fields. Consequently, even a small deviation from the $\bf{B \parallel  E}$ configuration gives rise to the additional contribution from conventional magnetoresistance (particularly visible in Figs. \ref{fig:F3}H,J). For two samples: $x=0.38$ (sample I) (Fig. \ref{fig:F3}J) and $x=0.4$ (Fig. \ref{fig:F3}L), namely with compositions closest to the band inversion point at low temperatures, the chiral anomaly is present even at room temperature. Moreover, for the latter sample ($x=0.4$) we report the reentrance of the chiral anomaly in the temperature range 275-310 K, preceded by its earlier disappearance at 170 K, as depicted in Fig. \ref{fig:F4}. The performed analysis provides a direct evidence that carrier concentration in the two mentioned samples is indeed sufficiently low to place the Fermi level within the nodal touching points, to which we will refer in the next paragraph. 

\begin{figure} [h]
	\includegraphics[width=1\textwidth]{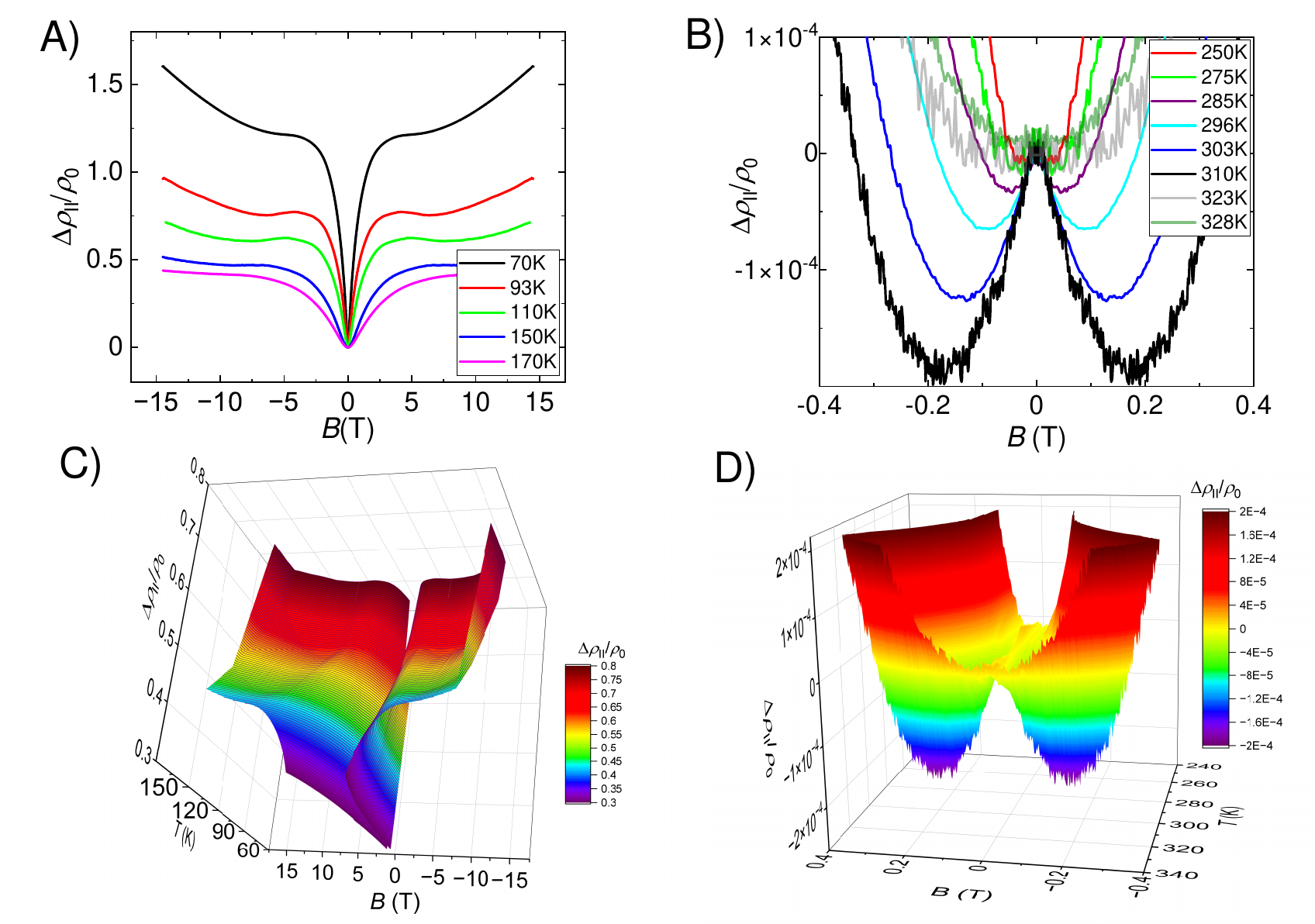}%
	\caption{ \label{fig:F4} Magnetic field and temperature dependence of the magnetoresistance  $\Delta\rho_{\parallel}/\rho_0$ for sample II in the parallel configuration of the magnetic and electric fields $\bf{B \parallel E}$. Chiral anomaly is observed for temperatures ranging from 93 K to 150 K (A). The chiral anomaly reentrance at temperatures 275-310 K is also observed (B).  Measurements were conducted using (A) 14.5 T magnet, (B)~0.6~T magnet. The depth of the negative magnetoresistance at 310 K equals 20 ppm and is two orders of magnitude lower than that at 110 K (1.1$\%$). (C) and (D) Three dimensional images, exhibiting changes of the depth of the magnetoresistance with temperature for panels (A) and (B), respectively. 
	}
\end{figure}

\begin{figure} [h]
	\includegraphics[width=1\textwidth]{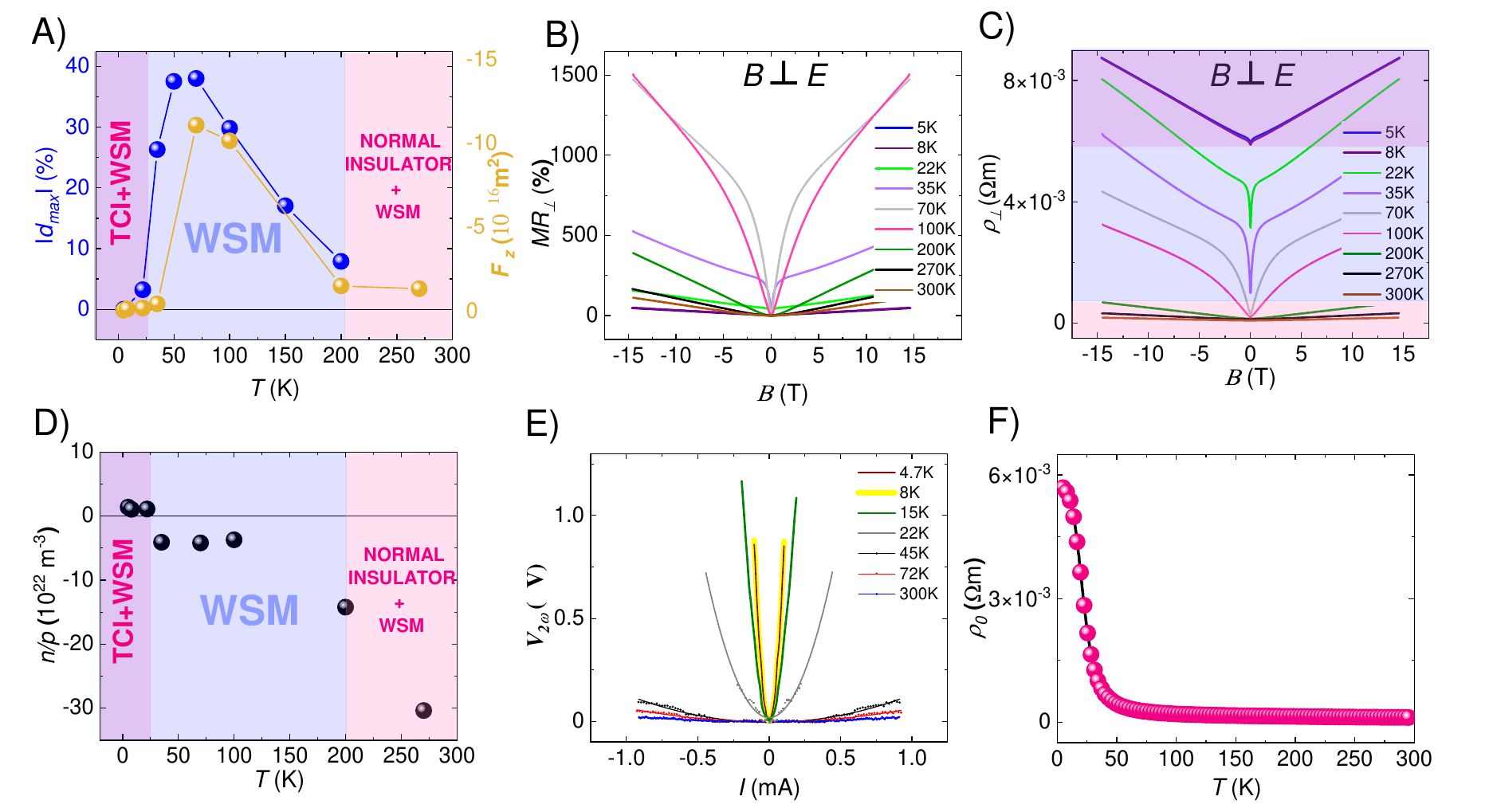}%
	\caption{\label{fig:F5} Magneto-transport results for sample I ($x=0.38$). (A) Temperature dependent maximum depth of the $MR_\parallel$  and the corresponding Berry curvature values. (B) Magnetic field dependence of the  $MR_\perp$ - magnetoresistance measured in the configuration with magnetic field perpendicular to the electric field. (C) Magnetic field dependence of the resistivity $\rho_\perp$ (for data from panel B). (D) Carrier concentration as a function of temperature from 5 to 300 K. (E) The quantum nonlinear Hall effect (QNHE) – the observed quadratic response of the second harmonic voltage to the applied AC current in the wide $T$ range: 4-300 K. (F) Zero field resistivity versus temperature illustrating TCI/WSM/NI phase transition.
	 }
\end{figure}

These observations are particularly intriguing when compared with the presence (and the magnitude) of the non-zero Berry curvature observed at these temperatures. Using the example of sample I we present a comparison of the temperature dependence of the depth of the negative contribution to magnetoresistivity $\rho_\parallel$ (the difference in $\%$ between the first maximum and the minimum of MR starting from the WAL region) and the Berry curvature (Fig. \ref{fig:F5}A). As already mentioned, the chiral anomaly for this sample is present in a wide temperature range, starting from 22 K up to 310 K as a result of a broaden region of very low carrier concentration $10^{16}-10^{17}$ cm$^{-3}$ (Fig. \ref{fig:F5}D). Indeed, the band gap extracted from the Arrhenius analysis in the temperature range from 75 K to RT is 3 meV.  This result supports our hypothesis, observed for all analyzed samples, that both the IAHE and the chiral anomaly emerge at very low carrier concentrations, i.e., when the Fermi level approaches the nearly closed energy gap. The highest values of the negative magnetoresistance contribution for sample I were obtained between 50 and 70 K which is consistent with the most prominent contribution of the extracted Berry curvature at these temperatures. This is also perfectly reflected by different magnetic field profiles of our $MR_\perp$. Figure \ref{fig:F5}B displays the magnetic field dependence of  $MR_\perp$  measured at $\bf{B \perp E}$. An unsaturated linear high-field  $MR_\perp$ is observed. We also obtain significant strengthening of magnetoresistance, especially at 70 K and 100~K, i.e in the same region where high anomalous Hall conductivity due to the Berry curvature was detected (depicted in Fig. \ref{fig:F5}A). Such a high magnetoresistance again confirms presence of the 3D WSM phase in our system since, as already mentioned, in contrast to the ordinary metals with parabolic or Kane-type non-parabolic dispersion relation, the momentum energy dependence of a 3D WSMs is linear. 

Figure \ref{fig:F5}C shows the magnetic field dependence of $\rho_\perp$ for sample I. In the qualitative manner we can distinguish three distinct regions which are intriguingly consistent with those depicted in Fig. \ref{fig:F5}A (for the chiral anomaly depth and for experimentally determined Berry curvature magnitude)\cite{wang_digging_2019}: 

- region (I) At lowest temperatures we observe small positive contribution to magnetoresistance, which can be assigned to WAL (see Supplementary Note S.IV). In the high-field range the linear MR is not surprising, since we have a semiconductor with zero band gap and the linear dispersion in energy versus momentum dependence.

- region (II) In the intermediate region the "WAL-like" behavior is very pronounced and broadens in the expected order with increasing temperature.

- region (III) In the highest temperature region only the linear MR is visible. This  region is a mixture of WSM phase with a heavy dose of the trivial phase, which is confirmed by the increase in the carrier concentration starting from 200 K up to RT (Fig. \ref{fig:F5}D). Therefore, the decrease of $\rho_\perp$ in this region does not surprise, as well as the suppression of both - chiral anomaly and the Berry curvature.

As can be seen from Fig. \ref{fig:F5}A,C,D, we have assigned the lowest temperature range to a mixture of phases: WSM and TCI. Indeed, we believe that a temperature-driven transition from WSM to TCI phase occurs in our samples which agrees with gradual changes in the magneto-transport behavior, while the WAL-like feature still persists. The emergence of the TCI phase is possible because topological surface states in TCIs are protected by (110) mirror-plane symmetry and are observed on high-symmetry crystallographic planes, namely (100), (110) and (111). As noted in {\bf Methods}, despite not being cleaved, our samples exhibit these crystallographic orientations, thus fulfilling the necessary condition for the emergence of  the topological surface states in a three dimensional TCI crystal.

Additional evidence for the presence of the TCI phase is provided by the observation of the quantum nonlinear Hall effect (QNHE) in this sample.  The QNHE is considered a hallmark of topological phases and its observation constitutes strong evidence for the existence of topological states in the material. Hence, this is as well present in Weyl semimetal phase as in TCI. Therefore the signal measured by us is a sum of two contributions. Figure \ref{fig:F5}E presents the results of the QNHE implying the presence of the Berry curvature dipole, reflected by the quadratic dependence of the second harmonic to the AC current excitation \cite{sodemann_quantum_2015,nishijima_ferroic_2023}. As can be seen, the most prominent signal is characteristic for TCI and in the WSM phase the signal is almost one order of magnitude smaller, however, all is within hundreds of nanovolts. In the region (III), i.e. where the trivial charge carriers dominate the QNHE almost completely vanishes.

The picture is complemented by a non-metallic zero-field resistivity $\rho_0$ versus temperature dependence (Fig. \ref{fig:F5} F), which can be interpreted as follows: the resistivity of the sample slowly increases as the temperature is lowered starting from RT, since as the sample undergoes the transition from the trivial to the WSM phase, the trivial bulk carriers are gradually "frozen out" and the metallic carriers, associated with surface states of 3D WSM remain. Their concentration is of the order of $10^{17}\,\mathrm{cm}^{-3}$. Below 50 K a sharp increase in $\rho_0$ is observed, indicating disappearance of the 3D WSM phase. From now on, only te metallic surface states of a TCI exist. The zero field resistivity of these states is high since the concentration of the surface-states charge carriers is very low ($10^{16}\,\mathrm{cm}^{-3}$). This interpretation is supported by the absence of other signatures of a 3D WSM, such as the chiral anomaly and the IAHE, accompanied by a drastic reduction in $\mathcal{F}$$_z$ at these temperatures. 

\begin{figure} [h]
	\includegraphics[width=1\textwidth]{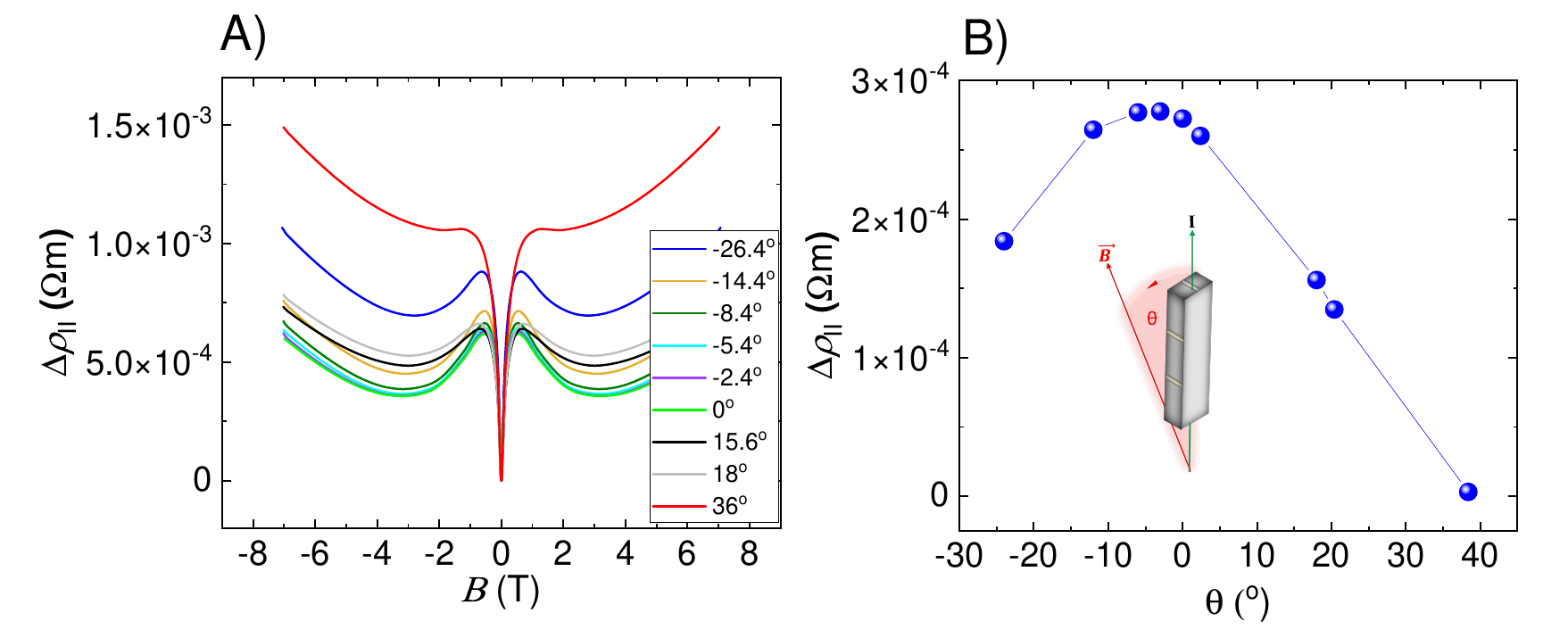}%
	\caption{\label{fig:F66} (A) Magnetoresistance $\Delta\rho_{\parallel}$ as a function of magnetic field $B$ for different angles $\theta$ between electric and magnetic fields.  (B) Angular-dependent $\Delta\rho_{\parallel}$  showing gradual suppression of negative contribution as the angle $\theta$ increases.
	 Since the negative magnetoresistance is most pronounced when the electric and magnetic fields are parallel, the presented results confirm the chiral anomaly-scenario. }
\end{figure}

Since chiral anomaly is not easy to identify and may be confused with the weak localization, long established in the literature \cite{bergmann_weak_1984}, we perform additional tests to verify these mechanisms.  In particular, we investigate the angular dependence of our negative contribution to resistivity $\Delta\rho_{\parallel}$, calculated from the inflection point to the minimum of $\rho_{\parallel}$. We observe that it vanishes already at small angles $\theta$ $\sim$ $40^\circ$ (Fig. \ref{fig:F66}). As the angle increases only the positive MR is present. Otherwise, if the observed effect were due to giant or colossal MR of magnetic origin, reported in literature \cite{taylor_resistivity_1968,ritchie_magnetic_2003,fert_nobel_2008} it should occur in both perpendicular and parallel MR configurations.  

We also exclude the geometry and size effects, the so called current-jetting, which can result in negative magneto-resistance, but is associated mainly with inhomogeneities of the electric field distribution \cite{reis_curr_jetting_2016,zhang_signatures_2016,liang_experimental_tests2018}. Additionally, for several samples we conduct magneto-transport measurements using the DC and AC techniques and we obtain the same results in both these approaches. All these efforts provide the evidence that chiral anomaly is the origin of the observed effects and the details of these studies are included in Supplementary Notes S.V - S.VII.

Moreover, we rule out the hypotheses regarding the origin of the linear MR in our samples arising from other reasons than Abrikosov's quantum limit \cite{abrikosov_quantum_1998,wang_linearMR_2012}, mentioned in the Introduction \cite{shen_revealing_2015}. In particular we exclude the conduction through two-component inhomogeneous system, such as PbSe matrix with embedded Ag metallic precipitates - the so called Parish-Littlewood model \cite{parish_classical_2005,hu_nonsaturating_2007}. As mentioned, the presence of Cr-Te nanoinclusions in our samples is very low and does not contribute to conduction. High compositional homogeneity of our samples is confirmed by the SEM/EDX measurements \cite{krolicka_cr_2025}. It is also confirmed by very similar resistivities obtained for both pairs of contacts. In this regard, we do not have a system containing multiple precipitates differing significantly in resistance that could affect the results. 

{\em Analysis of quantum oscillations in magnetoresistance and thermal conductivity}\\

During magneto-transport measurements, the quantum oscillations were observed for several of our samples, in a wide temperature range (Fig. \ref{fig:F3}B, D, F), providing evidence for good crystalline quality of these samples. Oscillation behavior is clearly visible in the configuration with aligned electric and magnetic fields, while in the perpendicular configuration, oscillations are not visible with the naked eye. This is a result of a large magnetoresistance $MR_\perp$ and to extract the oscillating component a second derivative of the raw data with respect to 1/B should be taken. The values of the obtained oscillation frequencies in both measurement configurations differ slightly. 

Quantum oscillations of magnetoresistance provide detailed and useful information on the Fermi surface of the system. Using the three terms in the Lifshitz-Kosevich formula \cite{ADAMS1959254,ROTH1966159,Shoenberg_1984}

\begin{equation}
	\begin{aligned}
	\frac{\Delta \rho}{\rho_0} \propto 
	\sqrt{B}
	\cdot exp\left(-\alpha \frac{T_D}{B}\right)
	\cdot \frac{X}{\sinh(X)}
	\cdot \cos\left[2\pi\left(\frac{Frq}{B} + \gamma\right)\right], \\[1em]
	\alpha = \frac{2\pi^2 k_B m_0}{\hbar e }=14.69 ~{\rm [T/K]}, 
	\quad
	X = \frac{\alpha T m^*}{B}, 
	\quad
	\gamma = \frac{1}{2} - \frac{\Phi_B}{2\pi}+\delta
	\end{aligned}
\end{equation}
one can extract the following key parameters: 
(1) the Dingle temperature $T_D$ and/or the quantum mobility $\mu_q = \frac{e \, \tau_q}{m^*}$ , where $\tau_q$ is the quantum lifetime, from the damping of the field dependent amplitudes of the oscillations, 
(2) the cyclotron effective mass $m^*$ from the temperature dependence of the oscillation amplitudes and 
(3) the oscillation frequencies $Frq_{max/min}$ that, according to the Onsager relation ($Frq = \frac{\hbar}{2\pi e} A$), correspond to the extreme cross-sections of the Fermi surface  $A_{max/min}$. The prefactor 	$\sqrt{B}$ in equation (1) is used for three dimenional topological systems and $\gamma$ is a phase factor, related to the Berry phase $\Phi_B$ and Maslov index $\delta$, to be discussed in more detail later.

\begin{figure} [h]
	\includegraphics[width=1.05\textwidth]{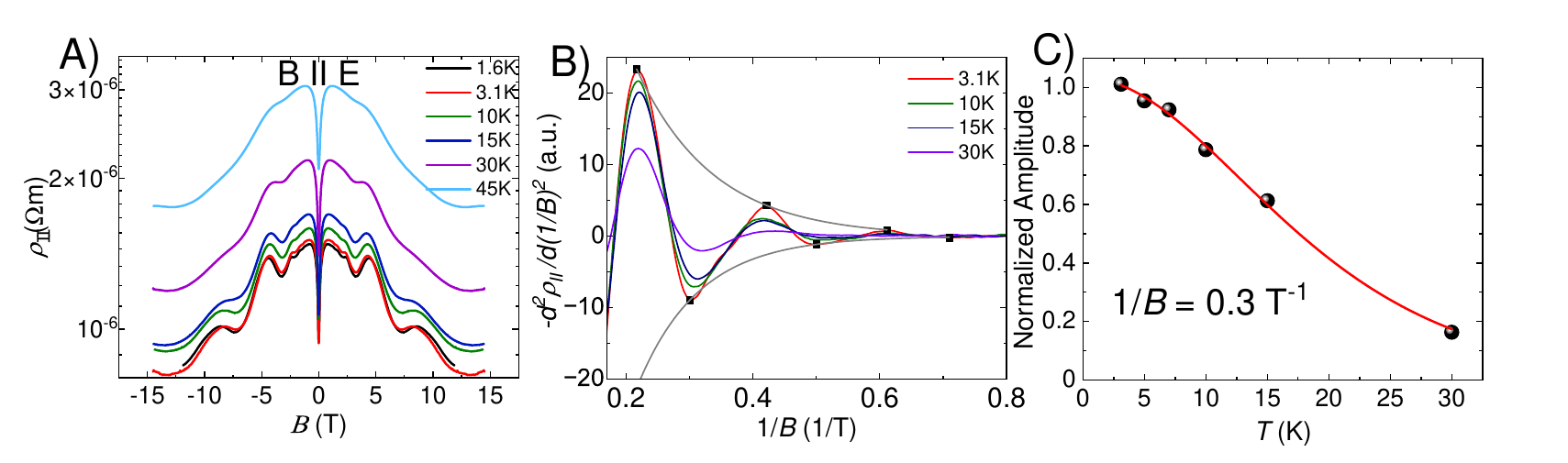}%
	\caption{\label{fig:F6} Analysis of the SdH oscillations for sample II ($x$=0.26) in the configuration with magnetic field applied parallel to the current. (A) Magnetic field dependent longitudinal resistivity at temperatures 1.6 - 45 K. (B) Shubnikov de Haas oscillations of the resistivity presented as a function of inverse magnetic field for selected temperatures from 1.6-30 K range. Background subtraction was obtained by the second derivative of $\rho_\parallel$ with respect to the inverse magnetic field. Grey solid line represents the envelope of the oscillations, reflecting the field-dependent decay of the oscillation amplitude (the Dingle damping) from which the quantum lifetime, quantum mobility and the Dingle temperature are determined. (C) Normalized to the lowest temperature value amplitudes of the quantum oscillations as a function of temperature, determined at a fixed value of the inverse magnetic field. Fitting the thermal damping term of  the Lifshitz-Kosevich formula (red solid line) enables extracting the effective mass of the charge carriers.	}
	\label{}
\end{figure}

 Here we present a detailed analysis of the oscillatory behavior of sample II ($x=0.26$). As shown in Fig. \ref{fig:F6}A, clear quantum oscillations for this sample are observed in a wide temperature range (1.6 - 45 K). We observe only a single oscillation frequency, most likely arising from the three-dimensional WSM states. According to the literature \cite{haldane_attachment_2014,potter_quantum_2014,Bulmash_2016}, SdH oscillations associated with two-dimensional fermions residing on Fermi arcs are expected to be observable only in samples with extremely small thicknesses, of the order of a few nanometers, where scattering effects remain sufficiently weak. In contrast, thicknesses of our samples are approximately 0.5-1 mm, meaning several orders of magnitude larger. Therefore, based on our data, we do not observe any signatures of Fermi arcs beyond the effects attributed to the chiral anomaly.

The values of the Fermi-surface parameters derived from the equation (1) are as follows: $\mu_q$ = 0.35  m$^{2}$/Vs,  $\tau_q$=44 fs and $T_D$ = 27 K, obtained from the fitting of the Dingle damping term to the oscillation amplitudes, as shown in Fig. \ref{fig:F6}B. The obtained value of the cyclotron effective mass $m^*$  from fitting the thermal damping term  is $(0.029 $+/-$ 0.003)$ $m_0$ (Fig.~\ref{fig:F6}C). The relatively small effective mass obtained in this analysis corresponds to carriers occupying linearly dispersing bands near the Weyl nodes. From the Heisenberg uncertainty principle $\Delta E \cdot \tau_q \geq 2\hbar$ we obtain the resulting width of the Landau level $\Delta E$ = 7 meV. Large width of the LL and short relaxation time together with high $T_D$, as compared i.e. with \cite{okazaki_extremely_MR_2026} suggest that we deal with some peculiarities related to redistribution of DOS in the magnetic field. The disorder-induced perturbations of the crystal structure, intentionally introduced to our material, likely results in the reduced Landau level separation and significant overlap between adjacent levels. Consequently, we may also deal with inelastic scattering (apart from elastic) and hence the reduced electron localization as compared with pure materials. Although quantum transport effects can still be detected, the disorder prevents the emergence of distinct, high amplitude SdH oscillations that are typically associated with high mobility Weyl fermions. 

\begin{figure} [htbp]
	\includegraphics[width=1\textwidth]{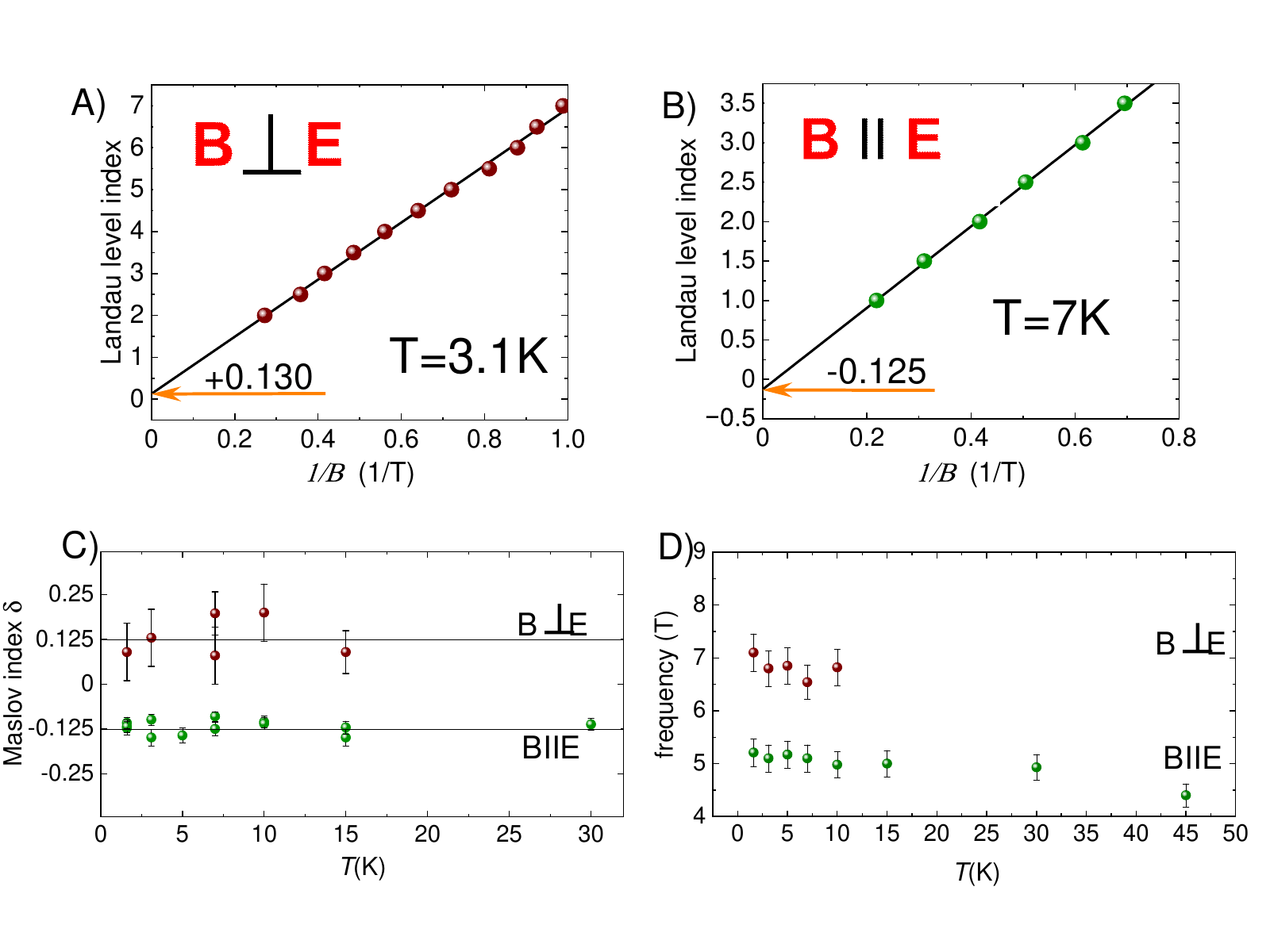}%
	\caption{ \label{fig:F7} Landau level index plots as a function of inverse magnetic field for sample II in two measurement configurations: with current applied perpendicular (A) and parallel (B) to the magnetic field enabling extraction of: (C) phase correction $\delta$ to the phase offset $\gamma$ of the oscillations corresponding to the intercept of the linear fit with the $y$-axis and (D) oscillation frequencies extracted from the slope of the linear fit.	}
\end{figure}

Quantum transport and especially SdH oscillations also bring useful information on the Berry phase \cite{PhysRevB.77.113407} via the expression for their phase factor:  $\gamma = \frac{1}{2} - \frac{\Phi_B}{2\pi}+\delta$. Berry phase is especially characteristic for topological systems: TIs and TCIs for which $\phi_B = \pi$. Additionally, Berry phase contains the phase offset/parameter $\delta$, called the Maslov index, which distinguishes 2D and 3D topological systems for which $\delta$ equals 0 and 0.125, respectively. Moreover, the sign of the Maslov index indicates whether the probed cross-section of the Fermi surface belongs to the maximum or minimum Fermi surface cross section \cite{Shoenberg_1984,hu__2016_TaP,orbanic_quantum_2017} . 

Landau level index analysis was performed by assigning integer indices to the oscillation extrema (Fig.~\ref{fig:F7}). For sample II ($x$=0.26) the obtained values deviate from those expected for trivial systems, supporting the presence of nontrivial band topology. As can be seen (Figs. \ref{fig:F6}B and \ref{fig:F7}A, B), the quantum limit for this sample is $B$ = 5 T ($1/B$ = 0.2 T$^{-1})$. The obtained Maslov indexes $\delta$ are: 0.130 and -0.125 for $B \perp E$ (Fig. \ref{fig:F7}A) and $B \parallel E$ (Fig.~\ref{fig:F7}B), respectively,  assuming the Berry phase exactly equals $\pi$ and the whole experimental error is included in $\delta$. This confirms that our system is a 3D Weyl semimetal. As already mentioned, the obtained $+/-$ signs indicate the maximum and minimum cross-sections of the Fermi surface, respectively. Figure \ref{fig:F7}C shows a comparison of $\delta$ values as a function of temperature for both measurement configurations. Small uncertainty of the obtained values indicates high reliability of the results. Also, the temperature dependent oscillation frequencies $Frq$ for both measurement configurations (Fig. \ref{fig:F7}D), obtained from the slope of the oscillation extrema, demonstrate very good reproducibility. $Frq_{max}$ of approx. 7~T and $Frq_{min}$ of approx. 5 T were obtained for the perpendicular and parallel configurations, respectively. These values reflect the maximal and minimal cross-sections of the Fermi surface, respectively. The calculated carrier density $n_{SdH}$ is 1.1 x $10^{17}$ cm$^{-3}$, while the Hall carrier density $n_H$ is 7 x $10^{17}$  cm$^{-3}$, which gives the ratio $n_H$ /  $n_{SdH}$ $\approx$ 6, instead of 8 (4~L-valleys x 2 = 8 Weyl cones), pointing to the ferroelectric distortion in one of the four $"L"$ valleys, which is in agreement with the literature \cite{liang_pressure-induced_2017,nishijima_ferroic_2023}. 

\begin{figure} [h]
	\includegraphics[width=1.1\textwidth]{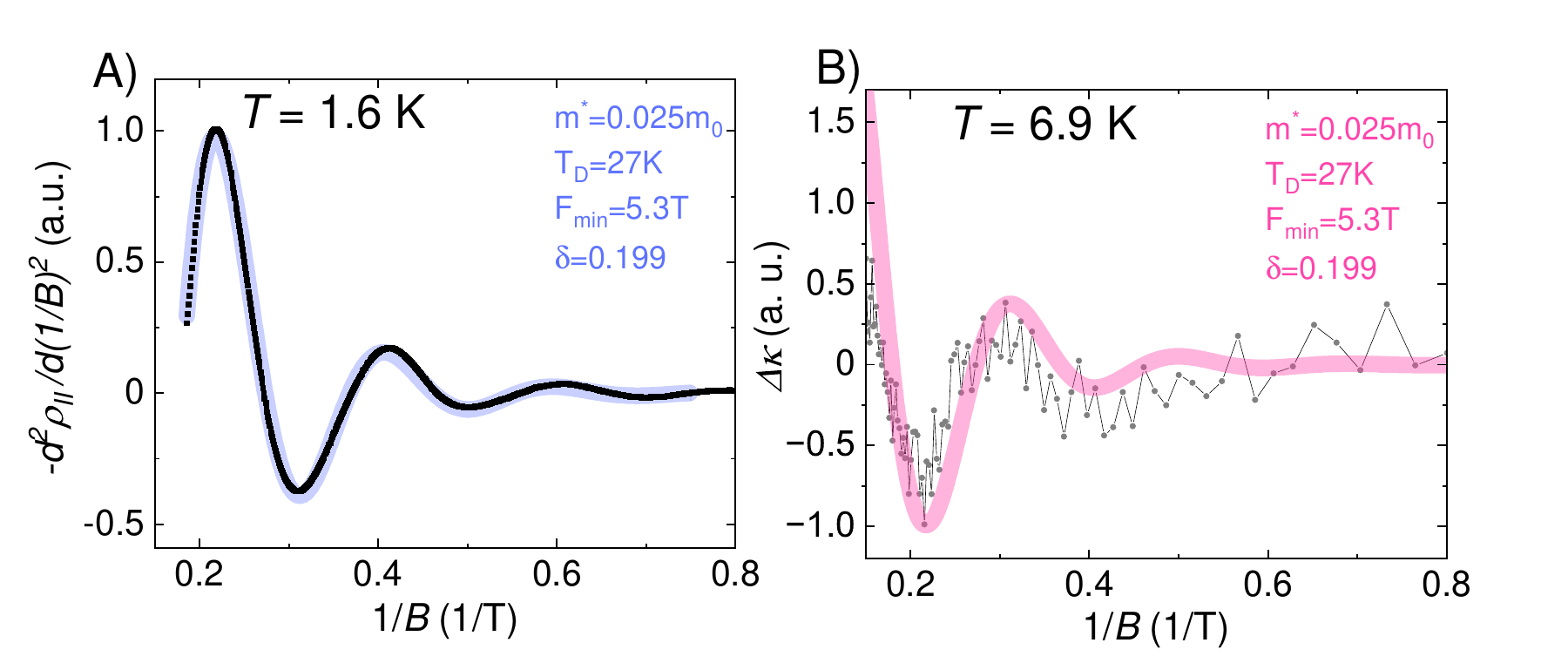}%
		\caption{ \label{fig:Foka}  Fitting of the Lifshitz-Kosevich formula using the determined material parameters to (A) the  second derivative of $\rho_\parallel$ with respect to the inverse magnetic field and (B) the thermal conductivity $\Delta\kappa$ as a function of the inverse magnetic field.}	
\end{figure}

To check the internal consistency of all parameters that we have determined from the SdH oscillations using the diagrammatic technique (each parameter, separately), we apply full Lifshitz-Kosevich formula to our data. Using previously determined values allows us to obtain a very good fit, as depicted in Fig. \ref{fig:Foka}A. Following the same procedure, we used the determined parameters and fitted them to the measured thermal conductivity data. The fit shows good qualitative agreement with the experimental data (Fig. \ref{fig:Foka}B).

\begin{figure}[htbp]
 	\includegraphics[width=1\textwidth]{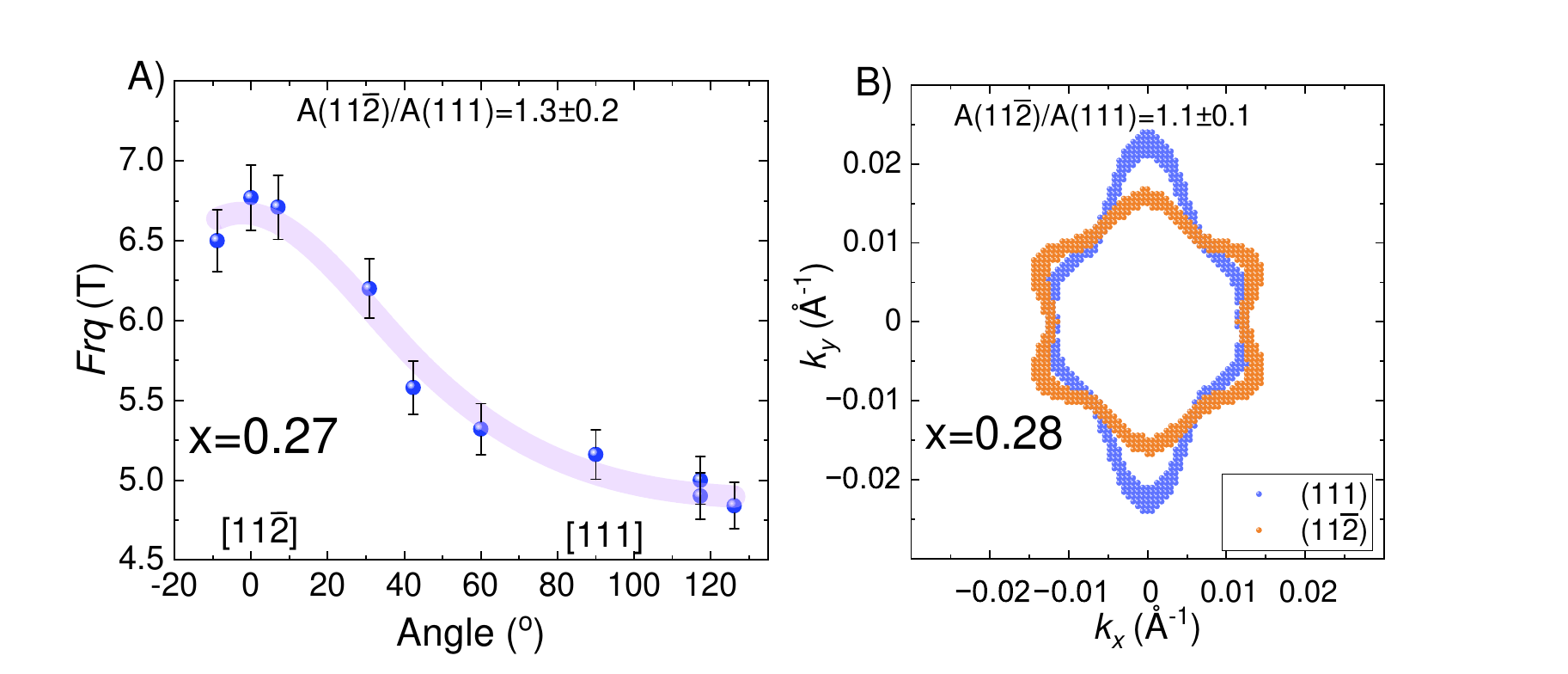}%
 	\caption{ (A) Angular evolution of the SdH oscillation frequency $Frq$$(\theta)$ for sample III ($x$=0.27) measured while rotating the magnetic field $B$ with respect to the normal of the surface plane. When the magnetic field is rotated within a symmetry plane of a single Fermi surface pocket, the frequency follows a trajectory that continuously connects extreme orbits corresponding to maximum and minimum cross-sectional areas. The local curvature of $Frq$($\theta$) reflects the geometry of the extreme orbit and allows one to distinguish between these two regimes. (B) The results of the theoretical calculations presenting two cross sections of the Fermi surface in $x$=0.28 \ce{Pb_{1-x}Sn_xTe} enclosed by the planes perpendicular to corresponding crystallographic directions [11$\bar{2}$] and [111] from panel (A).	}
 	\label{fig:F8}
 \end{figure}

To further probe the Fermi surface geometry, we perform the angle-dependent magneto-transport measurements by rotating the magnetic field over an angle of about 135 degrees in the plane perpendicular to the (111) crystallographic plane. We implement this procedure for sample III ($x$=0.27), thereby confirming the oscillation frequencies and the sign of the Maslov indexes, obtained for sample II (Fig. \ref{fig:F7}). The resulting measured angular dependence of the oscillation frequency for sample III is plotted in Fig. \ref{fig:F8}.
	
The $(111)$ crystallographic plane is a key plane for \ce{Pb_{1-x}Sn_xTe} crystals, since it is a high-symmetry plane in the rock-salt structure and the four equivalent "L" points, hosting electron valleys, lie along the $<111>$ directions\cite{burke_anisotropy_1970}. Since Fermi surface pockets are "pinned" to the "L" points, therefore they align with the four $<111>$ directions. Assuming this and the fact that we align the field with the principal axes of the Fermi surface cross-section, therefore by changing the angle $\theta$ between the magnetic field and one of the (111) crystallographic planes we can "scan" the Fermi surface cross-sections of one Fermi pocket. Since $Frq(\theta) \propto A_{extr}$, where $A_{extr}$ - extreme cross-sectional area perpendicular to the field direction, the angular dependent oscillation frequency is a function that connects maximum and minimum cross-sections of one valley. 

The tangent shape of this function reflects the sensitivity of the extreme orbits to the field rotation and the second derivative  $d^2F/d\theta^2$ describes their curvature. The sign of the second derivative is equivalent to the sign of the Maslov index, i.e. denotes whether the maximum or minimum cross section of the Fermi surface is probed. As \mbox{Fig.  \ref{fig:F8} } shows, the obtained results are the copy of those from Fig.  \ref{fig:F7}D, rotated by the angle of $90^o$. The frequencies $Frq_{max}$ = 6.8~T and $Frq_{min}$ = 5.2~T were obtained for the cross-sections $A_{max}$ and $A_{min}$, respectively. The ratio $Frq_{max}/Frq_{min} \propto A_{max}/A_{min} \propto 6.8/5.2 \propto 1.3\pm{0.2}$ indicates a nearly spherical shape of the Fermi surface and only slight anisotropy, being in agreement with Ref. \cite{liang_pressure-induced_2017}. Small anisotropy may, in our case, provide additional evidence that the Fermi level crosses the Fermi surface in the vicinity of the Weyl node touching points. This would be also consistent with the essentially single frequency observed in our quantum transport data. The extreme curvatures, i.e. the most concave and convex $Frq(\theta$) functions are at the angle values $\theta$ = 0$^{o}$ and $\theta = 90^o$, respectively, meaning they are separated by $90^o$ confirming the probed Fermi cross-sections are ellipsoids perpendicular to each other.

 Figure  \ref{fig:F8}A shows the oscillation frequency measured at different angles $\theta$ between the magnetic field and the normal to the sample plane. In Fig.\ref{fig:F8}B we show the calculated Fermi energy cross sections in planes (111) and (11$\bar{2}$), corresponding to the extreme frequencies of panel (A). We obtain $A(11\bar{2})/A(111)=1.1\pm 0.1$, which is in good agreement with the ratio of the extreme frequencies $Frq[11\bar{2}]/Frq[111]=6.8/5.2=1.3\pm 0.2$.   
 
\begin{figure}[htbp]
	\includegraphics[width=1\textwidth]{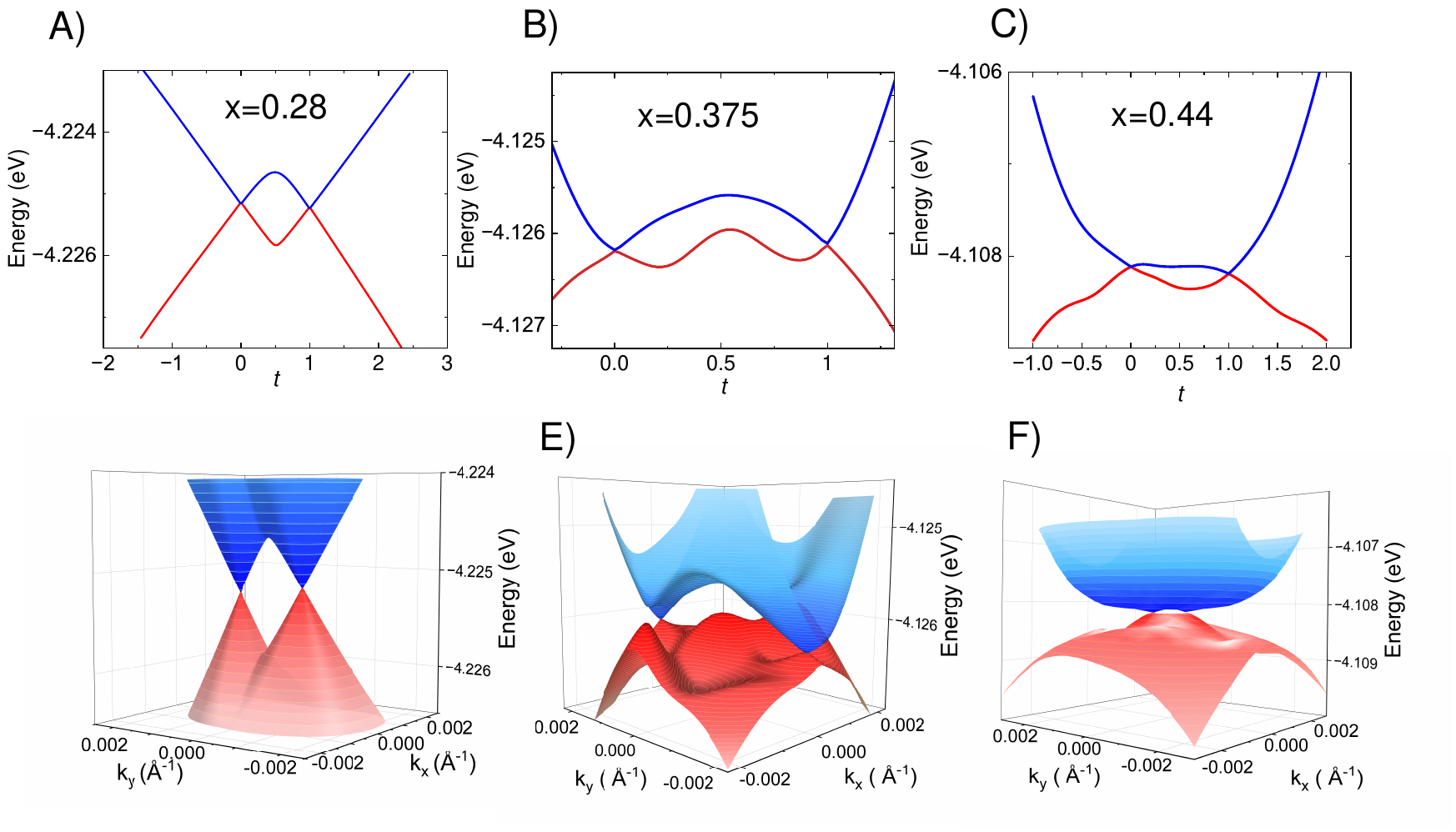}%
	\caption{ Dispersion relations for the lowest
		conduction and the highest valence bands in the vicinity of the Weyl
		points for $x=0.28$ (A), $x=0.375$ (B) and $x=0.44$ (C). The results are plotted in the properly chosen 
		coordination systems -- see the main text for explanation.   In panels (D), (E) and (F), the corresponding three dimensional dispersion
		relations are presented.  	}
	\label{fig:F11}
\end{figure}

Finally, it should be noticed, that for several samples at magnetic fields $B$ exceeding quantum limit, we observe also oscillations which are periodic in the magnetic field. One possible interpretation is that they correspond to the Aharonov-Bohm and Altshuler-Aronov-Spivak oscillations and originate from the quantum interference effects of charge carriers. The results for sample II are presented in more detail in Supplementary Note~ S.VIII.
\newpage
{\bf Theoretical calculations}\\
Our transport measurements findings (existence of the Weyl points in a broad temperature and crystal composition range and particular geometry of the Fermi surface cross-sections and their angular dependence) got support in the theoretical calculations based on
Density Functional Theory. In Table \ref{tab1} we present the determined topological charges in the reciprocal space with corresponding locations ($k_x$, $k_y$, $k_z$) for three representative Sn compositions from the expected WSM region. Figures \ref{fig:F11}A-C present dispersion relations for the
lowest conduction and the highest valence bands in 
the vicinity of the Weyl points for three different contents of Sn. Let us notice that all the plots are made in the coordination system with the $y$ axis along the line connecting two Weyl points and with the $x$ axis parallel to the (001) crystallographic plane. The dimensionless parameter $t$ in Figs.\ref{fig:F11}A-C describes the distance along the $y$ axis with $t=0$ and $t=1$, corresponding to the Weyl points of different topological charges. Panels D, E, F show three dimensional plots of the corresponding dispersion relations. As shown in Fig. \ref{fig:F11}, our theoretical calculations confirm the experimentally determined Sn composition range in which the WSM phase is present with sufficient accuracy: Weyl points with zero energy-gap are observed for Sn compositions in the range 0.28 $\leq x \leq$ 0.44 for comparison with the experimental Sn composition range: 0.25~$\leq~x~\leq$ 0.45. However, due to small distances between two Weyl points their experimental observation (e.g. Angular-Resolved Photoemission Spectroscopy) is impossible at present.\\

\begin{table}
	\caption{\label{tab1}}Sn concentration x, positions of the Weyl points
	($k_{x,y,z}$ in \angstrom) in the
	Brillouin Zone for 64  atom supercell,   topological charge
 for \pst. 
	\begin{ruledtabular}
		\begin{tabular}{cccccc}
			x&$k_x$&$k_y$&$k_z$&Charge\\
			0.281&4.102e-05&-8.203e-4& 4.666e-3&+1\\
			0.281&2.519e-4&6.953e-4&4.727e-3&-1\\
			0.375& -1.531e-3&-1.015e-3&7.0312e-4&+1\\
			0.375& 1.546e-3&1.515e-3&7.3437e-4&-1\\
			0.438&0.0011&-9.5e-4&1.475e-3&+1\\
			0.438&0.00035&1.225e-3&1.1e-3&-1\\
		\end{tabular}
	\end{ruledtabular}
\end{table}

{\bf Conclusions}\\
\ce{Pb_{1-x}Sn_xTe}:Cr substitutional alloy with Sn content in the window $0.25<x<0.45$ constitute new intriguing Weyl semimetal system originating from the influence of the chemical alloy disorder on the multivalley electronic structure of this semiconductor, resulting in the sequential band inversions at zero energy gap in various valleys. Our magnetotransport experiments support our theoretical calculations that identified the pairs of Weyl nodes with topological charges.

\ce{Pb_{1-x}Sn_xTe}:Cr alloys exhibit several peculiar magneto-transport phenomena: including the nonlinear quantum Hall effect, the intrinsic anomalous Hall effect and the chiral anomaly under parallel electric and magnetic fields. These effects are believed to originate from the nontrivial Berry curvature and the Berry monopoles near the Weyl points, providing an intrinsic contribution to their emergence.

In this study, we demonstrate magnetotransport evidence for the presence of the 3D Weyl semimetal phase in \ce{Pb_{1-x}Sn_xTe}:Cr  bulk crystals and verify the relationship between the magnitude of the determined Berry curvature and the electrical properties of these materials. To investigate these effects, we achieve pinning of the Fermi level in the vicinity of the nodal touching points in \ce{Pb_{1-x}Sn_xTe}  over a wide range of Sn compositions by doping crystals with Cr. We obtain a consistent description of the chiral anomaly in terms of experimentally estimated average Berry curvature and its temperature and composition dependencies. For several samples chiral anomaly covers almost the entire temperature range. Moreover, for one of our samples, we observe the striking phenomenon of the room temperature re-entrance of the chiral anomaly, in addition to its presence in the 93–150~K temperature range. Additional insight into this issue is provided by the analysis of the SdH oscillations, which reach the quantum limit at relatively low magnetic field of approximately 5~T. The finite oscillation frequency observed at an angle of 90 degrees, provides direct evidence that the oscillations do not originate from a 2D electron gas, since in that case, no oscillations would be expected for the $\bf{B \parallel E}$ configuration. Thus, the angular dependence of the SdH frequency, together with the $\pi$ Berry phase extracted from the Landau level index plot, confirms the 3D origin of the electronic states and points to a topological WSM phase.  Moreover, additional evidence for the 3D character of the Fermi surface, is provided by the Maslov index of 1/8, extracted from the SdH oscillations with high accuracy. Further insight from the quantum oscillation measurements is provided by the analysis of the Fermi surface cross-sections measured in two perpendicular configurations. We determine an anisotropy factor of 1.3 $\pm$ 0.2, in contrast to the much larger values of 2.4-3.2, expected for the ellipsoidal Fermi surfaces shape of parent PbTe, SnTe compounds. These results are also in agreement with our first-principles theoretical model.
In conclusion, we believe that our experimental observations and analysis of variety of magnetotransport effects with the Berry curvature contributions to both resistivity and Hall effect, strongly support the presence of the Weyl semimetal phase in \ce{Pb_{1-x}Sn_xTe}:Cr. In the presence of chemical (electronic) alloy disorder the band inversion at zero-gap state takes place sequentially, discriminating the four electron valleys. To reach the physical regime with the sizeable topological contributions to electron transport we successfully applied doping with the resonant donor centre of Cr pinning the Fermi level close to zero-gap state.\\
\newpage
{\bf \large Methods}

{\bf Growth and morphology}

The Bridgman method was employed to grow bulk  \ce{Pb_{1-x}Sn_xTe}:Cr ingots with large single-crystalline grains, covering a full range of Sn nominal composition (0 $\leq x \leq$ 1) and heavily doped with Cr - up to 2 at.\%. However, we note that, due to strong segregation, the studied samples are expected to exhibit the regime where Cr constitutes up to 0.5 at.\% of the cations. All the details regarding characterization of the obtained crystals, including chemical composition, morphology and crystal structure - namely scanning electron microscopy (SEM)  together with the energy dispersive X-ray spectrometry (EDX), X-ray diffraction (XRD), secondary ion mass spectrometry (SIMS) and energy-dispersive X-ray fluorescence (EDXF) - have been reported in our previous publications \cite{krolicka_cr_2025,gas_magnetic_2021}. Also our magnetic measurements, including SQUID magnetometry and Electron Paramagnetic Resonance (EPR), performed during previous studies \cite{story_transport_1992,story_pbsecr_1995,krolicka_cr_2025,gas_magnetic_2021} indicate that host material is paramagnetic and contains various ferromagnetic Cr-Te nanoinclusions embedded in the \ce{Pb_{1-x}Sn_xTe}:Cr matrix. Nevertheless, their amount is relatively small and hence they do not affect transport of our samples. Samples for transport measurements were prepared by cutting rectangular parallelepiped specimens from the ingots perpendicular to the growth direction.
For three most representative samples the crystal orientation was determined using the Laue-X-ray diffraction method.  According to the Laue diffraction analysis, the surface of sample~I with composition $\ce{Pb_{1-x}Sn_{x}Te}$:Cr, ($x$=0.38) consists of two grains with the (111) crystallographic orientation and a smaller region with the (110) orientation. Samples II and III, subsequently employed for angular-dependent transport measurements, expose the following crystallographic planes: sample II with composition $\ce{Pb_{1-x}Sn_{x}Te}$:Cr, ($x$=0.26) possesses a surface corresponding to (111) crystallographic plane, while the perpendicular side facets correspond to (110) and (112) planes; sample III, in contrast, with composition $\ce{Pb_{1-x}Sn_{x}Te}$:Cr, ($x$=0.27), was cut from sample II and has the crystallographic planes rotated by 90 degrees relative to those of sample II.

{\bf Experimental}\\
Electrical transport measurements were conducted to examine the existence of a Berry phase and to establish the emergence of a three-dimensional Weyl semimetal (3D WSM) phase in \ce{Pb_{1-x}Sn_xTe}:Cr (0 $\leq x\leq$1) crystals. Samples were prepared in a conventional Hall-bar geometry, and indium contacts were soldered to ensure its stability and low-resistance. The measurements were performed using two pairs of voltage contacts,  placed opposite each other, and the small discrepancy of the results collected from both pairs (less than 5\% difference in the determined values of carrier concentration) indicates a high degree of homogeneity in the grown crystals, as expected for system doped with a resonant Cr donor. 
Measurements were performed in a cryogenic system equipped with a superconducting magnet providing magnetic fields up to 14.5 T. The temperature was regulated within the range 1.5–310~K using two Lake Shore Temperature controllers: with one sensor mounted in close proximity to the sample and the second located in the variable temperature insert (VTI). This temperature control system allowed an accuracy better than 0.01 K.
Transport signals were measured using a digital lock-in amplifiers (MFLI 500 kHz, Zurich Instruments), enabling high-sensitivity detection of both AC and DC voltage and current components. Complementary DC measurements were carried out using a Keithley current source in combination with a set of Keithley nanovoltmeters. This allowed to verify the compatibility between these two techniques and to identify potential similarities or differences in the manifestation of the chiral anomaly transport under AC and DC excitation conditions.

{\em Transverse configuration} ($\bf{B \perp  E})$

Magneto-transport measurements of the resistivity tensor were first performed in magnetic fields up to 14.5 T, applied perpendicular to the electric current 
($\bf{B \perp E}$). The observation of the intrinsic anomalous Hall effect (IAHE) was of significant importance. The emergence of such a contribution, independent of external magnetic field, was treated as an initial indication of a finite three-dimensional Berry curvature and, consequently, of the realization of a 3D WSM phase in the investigated samples.

{\em Longitudinal configuration} ($\bf{B \parallel E})$

Subsequently, magneto-transport measurements were performed with the magnetic field applied 
parallel to the electric field ($\bf{B \parallel E}$), again up to 14.5 T. This geometry was chosen to measure negative magnetoresistance, commonly associated with the chiral anomaly in Weyl semimetals.

{\em Quantum nonlinear Hall effect}

The measurements of the second-harmonic Hall response as a function of the applied AC current were performed under zero external magnetic field. A quadratic scaling of the second-harmonic transverse voltage with the applied current was analyzed as a characteristic signature of the quantum nonlinear Hall effect. Observation of this effect provided complementary evidence for nontrivial Berry curvature and supported the identification of the Weyl semimetal phase in \ce{Pb_{1-x}Sn_xTe}:Cr.

{\em Shubnikov-de Haas oscillations}

An analysis of the Shubnikov-de Haas oscillations was performed for two selected samples, including the angular dependence on the magnetic field $B$. The analysis provided the important material parameters, such as effective mass, Dingle temperature and relaxation time. Moreover, the oscillation frequencies obtained in both measurement configurations allowed the determination of the extreme Fermi surface cross-sections. 

{\em Thermal conductivity measurements}\\
Thermal-conductivity measurements were carried out using the absolute axial heat flow method in a Quantum Design PPMS equipped with a 14 T superconducting magnet. A sample was attached to a custom-made sterling-silver holder using DuPont 4929 N silver paste, and the temperature gradient was induced using a 5 k$\Omega$ Micro-Measurements strain gauge (used as a heater) mounted at the opposite end of the sample. The base temperature was measured using a Cernox thermometer, whereas the temperature differences along the sample and between the holder and the heater were determined using Constantan - Chromel thermocouples calibrated as a function of magnetic field \cite{matusiak_practical_2026}.

{\em Technical details of calculations}

For the density functional theory calculations we use open-source OpenMX package \cite{openmx}
with the fully relativistic pseudopotentials (version 2019). In order
to obtain proper energy gap and proper symmetry of the bands for PbTe,
the spin-orbit strength for $6p$ Pb orbitals was reduced to 0.554,
comparing to the original strength. The necessity of such reduction
was previously discussed in Ref. \onlinecite{lusakowski_calculated_2011}.
Infinite \pst\ random mixed crystals are modeled by $2\times 2 \times
2$ supercells containing 64 atoms (32 cations and 32 anions). The
lattice parameter,   
\begin{equation}
	a = 6.46-0.16x\ \ \angstrom
\end{equation}
was assumed as linearly interpolated between lattice parameters for PbTe
(6.46~\angstrom) and SnTe (6.30~\angstrom). All samples were geometrically optimized with the criterion for
forces $5\times~10^{-4}$~Ha/Bohr.
After calculations OpenMX provides set of the tight binding parameters
which enables to perform other necessary calculations -- searching for
the Weyl points, their charges and sections of the Fermi surfaces. 
The method of finding the Weyl points was described in
Ref. \onlinecite{lusakowski_band_2021}. After approximate localization of WP in certain three
dimensional cubic k-space we perform additional calculations to
find more accurately its k-coordinates and the energy gap.
It should be noticed here that,
first, as was pointed out in
Refs. \onlinecite{lusakowski_alloy_2018,lusakowski_band_2021}, the electronic band structure strongly
depends on the spatial distribution of Sn atoms in the
supercell. Usually we obtain results with positive or negative energy
gaps. That is why finding the system containing Weyl points is not an easy
task. Probably the problem would be partially solved by considering
larger supercells, but the calculations are long-lasting and this approach
is impractical.
Secondly, as previously in Ref. \onlinecite{lusakowski_band_2021}, the separation of the
Weyl points in k-space is very small, thus it is impossible to observe experimentally
Fermi arcs in the Angular Resolved Photoemission Spectroscopy experiments.

Interpretation of Shubnikov-de Haas measurements requires knowledge of
Fermi surface cross section areas. In calculations we focus on
sections for which the normal to
the section is along [111] and [11$\bar{2}$] crystallographic
directions.
We use the following method: we start with the set of equaly distributed
points ($k_x,k_y,k_z=0$) in (001) plane; next we perform proper
rotations of this plane 
such that the normal to the rotated plane is along [111] or
[11$\bar{2}$] direction, respectively; then vector ($k_x,k_y,k_z=0$) becomes
($k_x',k_y',k_z'$); for this vector we calculate the energy of state in the first
conduction band; after the band structure calculations we perform
inverse rotation to the (001) plane; finally we choose Fermi energy and
for this Fermi energy we obtain set of ($k_x, k_y$) points which may be
plotted on two dimensional plane and the area of the resulting figure
may be calculated.

\begin{acknowledgments}
	
	This study has been supported by the National Science Centre (Poland) through Project OPUS (UMO2017/27/B/ST3/02470) and by the National Science Centre for Development (Poland) through Grant TERMOD No. TECHMATSTRATEG2/408569/5/NCBR/2019 and by the Foundation for Polish Science project “MagTop” No. FENG.02.01-IP.05–0028/23 cofinanced by the European Union from the funds of Priority 2 of the European Funds for a Smart Economy Program 2021–2027 (FENG). M.M. acknowledges financial support from the National Science Centre (Poland) under Research Grant No. 2025/59/B/ST3/00232

\end{acknowledgments}


%

\clearpage

\setcounter{page}{1}
\renewcommand{\thepage}{S\arabic{page}}

\setcounter{equation}{0}
\renewcommand{\theequation}{S\arabic{equation}}

\setcounter{figure}{0}
\renewcommand{\thefigure}{S\arabic{figure}}

\setcounter{table}{0}
\renewcommand{\thetable}{S\arabic{table}}

\setcounter{section}{0}
\renewcommand{\thesection}{S.\Roman{section}}


\onecolumngrid
\begin{center}
	{\Large\bfseries Supplemental Information for}\\[6pt]
	{\large Tunable chiral anomaly in electron magnetotransport in the Weyl semimetallic \ce{Pb_{1-x}Sn_xTe}:Cr alloy }
\end{center}

\vspace{1cm}

{\large \bf CONTENTS}\\
S.I. Experimental estimation of the Berry curvature \hfill S2\\
S.II. Fitting the two-carrier Drude-Lorentz model \hfill S4\\
S.III. Maximal magnetoresistance \hfill S5\\
S.IV. Fitting Weak antilocalization and chiral anomaly contributions \hfill S7\\
S.V. Comparison of the magnetic field-dependent resistance measured in two
configurations \hfill S9\\
S.VI. Absence of current jetting \hfill S10\\
S.VII. Comparison of DC and AC measurement techniques \hfill S11\\
S.VIII. The Aharonov-Bohm and Altshuler-Aronov-Spivak oscillations \hfill S12\\
References \hfill S13\\

\newpage
\section{Experimental estimation of the Berry curvature }

The magnetic field dependences of the resistivity tensor components $\rho_{xx}$ and $\rho_{yx}$ are presented in the main text. The Berry curvature may be obtained from the analysis of the total Hall conductivity ($\sigma_{yx}$), which  is the sum of the conventional Hall ($\sigma_N$) and anomalous Hall contribution ($\sigma_{AHE}$), respectively \cite{jungwirth_anomalous_2002,ong_lee,liang_pressure-induced_2017}:
\begin{equation}
	\begin{aligned}
		\\\sigma_{yx} = \sigma_N + \sigma_{AHE}
		.
	\end{aligned}
\end{equation} 

$\sigma_{AHE}$ is related to the Berry curvature ($\Omega_z$) via the equation:
\begin{equation}
	\begin{aligned}
		\\\sigma_{AHE} = 
		\frac{e^2}{\hbar} 
		\int
		\frac{{d}\vec{k}}{(2\pi)^3}
		\cdot f_k
		\cdot \Omega_z (\vec{k}) =
		\frac{e^2}{\hbar}
		\cdot n
		\cdot \mathcal{F}_z ,
	\end{aligned}
\end{equation} 
where
\begin{equation}
	\begin{aligned}
		\\\mathcal{F}_z =  
		\int
		\frac{{d}\vec{k} f_k \Omega_z (\vec{k})}{(2\pi)^3 n } ,
	\end{aligned}
\end{equation} 
where $e$ is the electron charge, $\hbar$ - the reduced Planck constant, $\Omega_z$ - the Berry curvature, $f_k$ is the Fermi-Dirac distribution, $\mathcal{F}_z$ - the Berry curvature averaged over the Fermi surface.

We extract the anomalous Hall resistivity contribution $\rho_{yx}$ from the linear extrapolation of the high-field Hall resistivity curve (Fig. 2b in the main text). We assume that the intrinsic contribution is constant in the magnetic field. Since we are interested in determining  the intrinsic (magnetic field-independent) contribution to $\sigma_{AHE}$, we can also take the zero field value of  $\sigma_{xx}$. Then $\sigma_{AHE}$  can be calculated using the equation:

\begin{equation}
	\begin{aligned}
		\\\sigma_{AHE} =  
		\frac{\rho^{AHE}_{yx}}{({\rho^{AHE}_{yx}})^2 + \rho_{xx}^2}
	\end{aligned}
\end{equation} 
Finally, we estimate the value of the averaged Berry curvature $\mathcal{F}_z$ using the equation S5:
\begin{equation}
	\begin{aligned}
		\\\mathcal{F}_z =  
		\frac{\hbar  }{{e^2}} 
		\cdot	\frac{\sigma_{AHE} }{{ n}} 
	\end{aligned}
\end{equation} 
Figure S1 shows the averaged Berry curvature $\mathcal{F}_z$ as a function of temperature for several representative Pb$_{1-x}$Sn$_x$Cr$_y$Te samples (0.26$\le$x$\le$0.42, 0.01$\le$y$\le$0.02), i.e. from the WSM regime.

\begin{figure} [htbp]
	\vspace*{-1.5cm}
	\includegraphics[width=0.94\textwidth]{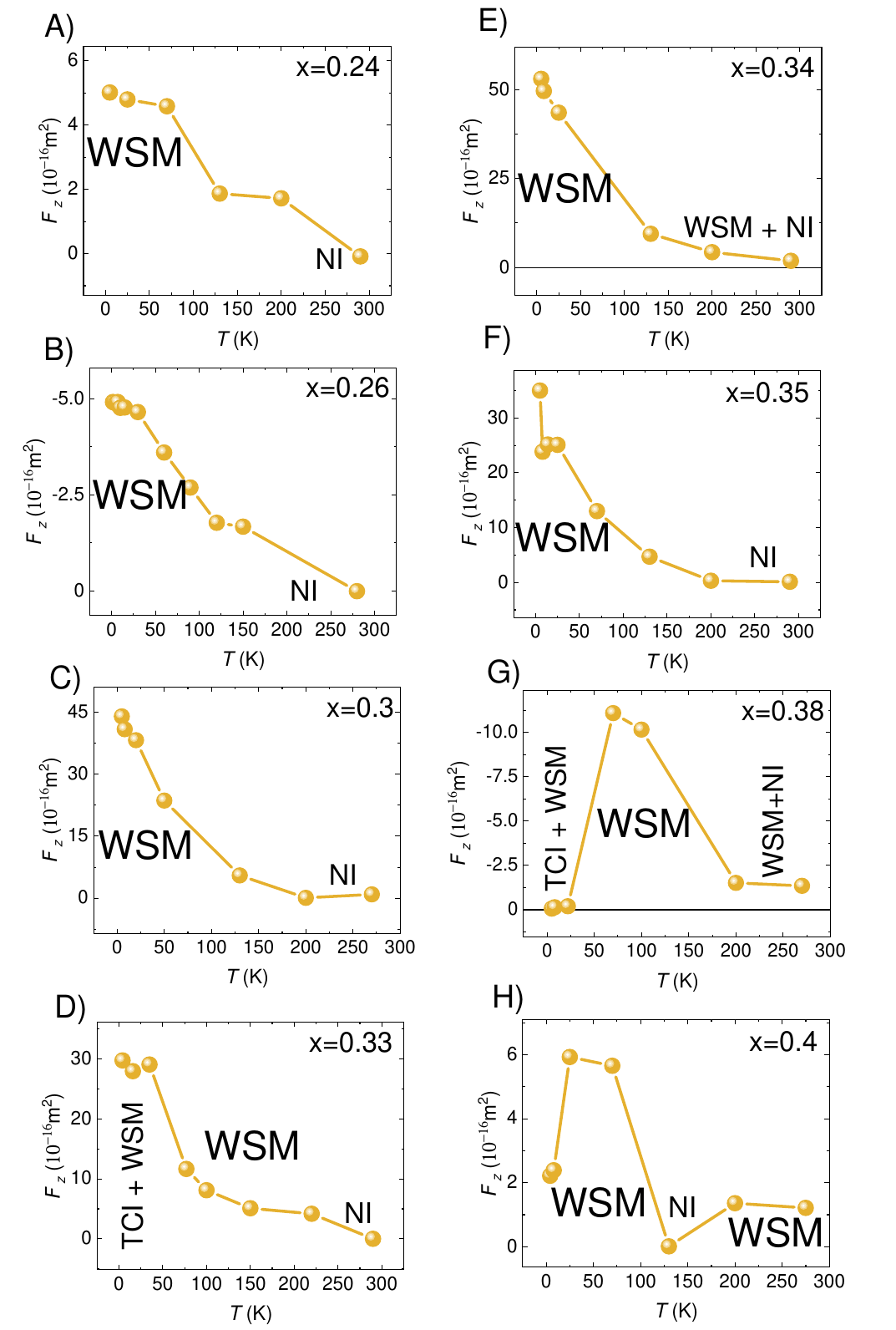}%
	\vskip -2ex
	\caption{\label{fig:Berry} Berry curvature $\mathcal{F}$$_z$ versus temperature for all samples from the WSM region ($0.25<x<0.45$) with the corresponding dominant phases (WSM, TCI, normal).  }
	
\end{figure}
\section{Fitting the two-carrier Drude-Lorentz model  }
To support the hypothesis of single-carrier conductivity in our samples we have also attempted to fit a two-band conduction model to our $\rho_{xx} (B)$ and $\rho_{yx} (B)$ data. When the magnetic field increases to 14.5 T, the positive $\rho_{xx}$ does not show any signature of saturation. In contrast to the simulated $\rho_{xx}$ using the two-carrier model along with the carrier concentration and mobility extracted from the Hall resistivity curves, the measured $\rho_{xx}$ and $\rho_{yx}$ data are far deviated from the two-carrier model. This means that a notable mechanism dominates the magnetotransport in the current system (despite the low magnetic field regime, in which single-carrier conduction occurs). We assign this dominating mechanism to IAHE associated with large Berry curvature arising when the system undergoes inversion symmetry breakdown.
\newpage

\section{Maximal magnetoresistance  }
\begin{figure} [htbp]
	\vspace*{-1.2cm}
	\includegraphics[width=1.1\textwidth]{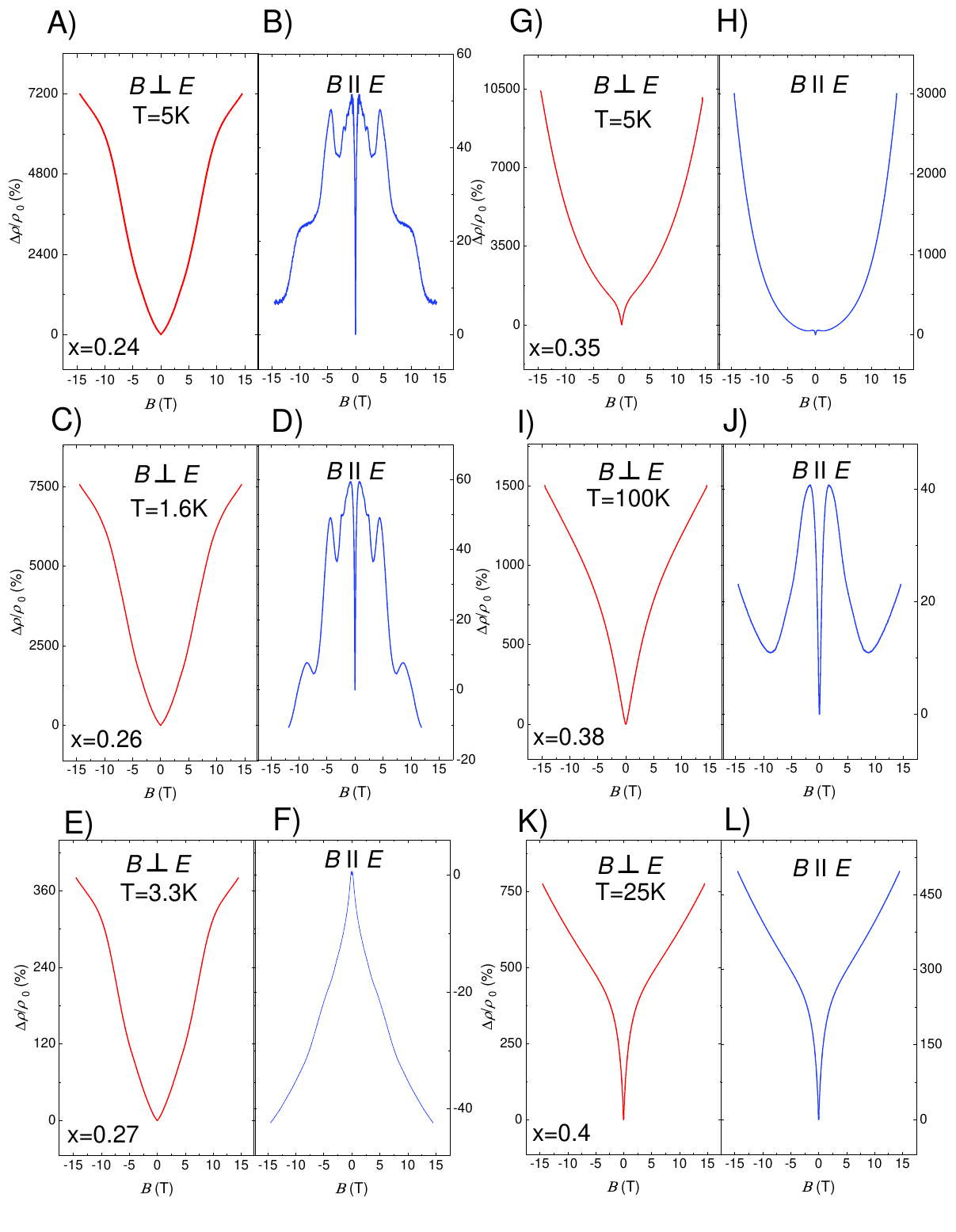}%
	\vskip -6ex
	\caption{\label{fig:MR} Maximal $\rho/\rho_0$ versus $B$ values for all WSM samples, presented in two measurement configurations: ${B}\perp{E}$ (red) and ${B}\parallel{E}$ (blue).  }
\end{figure}

The comparison of perpendicular and parallel magnetoresistance versus magnetic field for samples from the WSM region is plotted in Fig. \ref{fig:MR}. Presented curves show data for temperatures where the maximal magnetoresistance was detected for each sample.

\newpage

\section{Fitting Weak antilocalization and chiral anomaly contributions }

For sample II (Fig. 3D in the main text) we perform the fitting analysis of our $\rho_\parallel$ data for several representative temperatures. We utilize the following formula:

\begin{equation}
	\begin{aligned}
		\Delta\sigma_{\parallel}(B) =
		4C_wB^2-C_{WAL}  \left(\sqrt{B}
		\frac{ B^2 }{{B^2+B_c^2}} 
		+\gamma B^2	
		\frac{ B_c^2 }{{B^2+B_c^2}} \right)
	\end{aligned}
\end{equation} 
where: $C_w$ - chiral anomaly factor, $C_{WAL}$ - weak antilocalization factor, $B_c$ - critical field, where WAL dependence changes from $-B^2$ to $-\sqrt{B}$, $\gamma$ - proportionality factor of the above components. 
The above formula consists of two contributions: (i) chiral anomaly proportional to $B^2$ term \cite{son_chiral_2013} and (ii) quantum interference term in the form of the weak antilocalization for the 3D WSM system \cite{lu_weak_2015,zhang_signatures_2016}. \cblu{The prefactor 4 before $C_w$ accounts for the valley degeneracy. In contrast, the corresponding prefactor in the WAL term is equal to 1, because in the presence of elastic inter-valley scattering the valleys do not contribute independently to conductivity but rather as a single coherent system, even for systems with multivalley degeneracy \cite{fukuyama_non-metallic_1980}.}

\begin{figure} [htbp]
	\vspace*{-2.5cm}
	\includegraphics[width=0.9\textwidth]{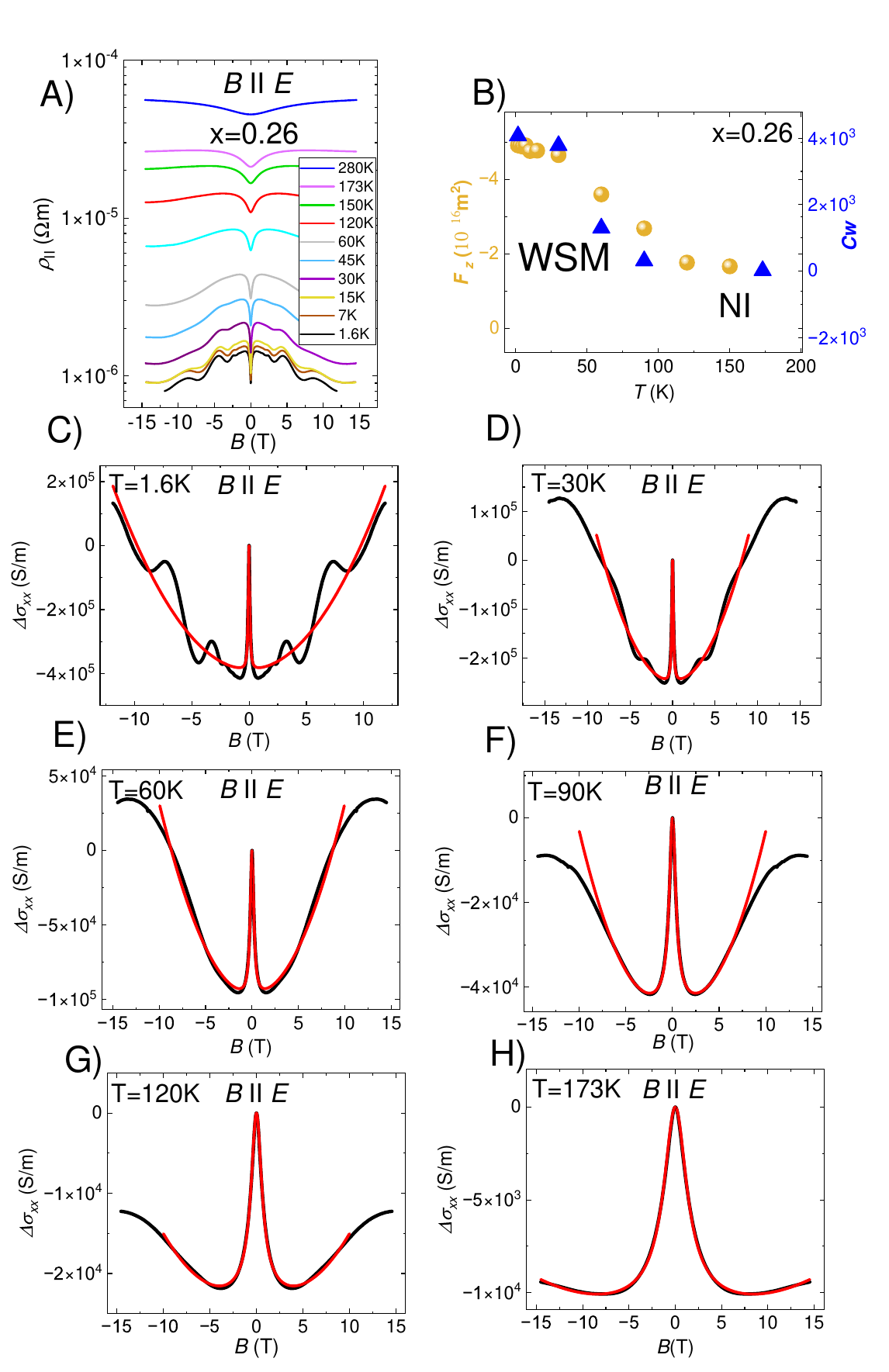}%
	\vspace*{-0.86cm}
	\caption{\label{fig:Fig3D} \cblu{ (A) Resistivity $\rho_\parallel$ versus magnetic field in the configuration with aligned electric and magnetic fields (${B}\parallel{E}$). Copy of panel (D) from Fig. 3 (main text) for the convenient comparison with the conductivity fits (panels C-H). (B) Comparison of the chiral anomaly factor ($C_w$) (blue triangles) with the calculated Berry curvature ($\mathcal{F}_z$)(golden spheres) for this sample. (C-H) Fitting results (red) for the experimental magnetoconductivity data (black) at several representative temperatures, according to the formula S6. }  }
\end{figure}
As can be seen from Fig. \ref{fig:Fig3D} we achieved a very good description of the experimental data. Deviation of the high-field conductivity from the theoretical model originates from a small admixture of large positive magnetoresistance due to a small misalignment between the electric and magnetic fields.

\newpage
\section{Comparison of the magnetic field-dependent resistance measured in two configurations }
\begin{figure} [htbp]
	\includegraphics[width=1\textwidth]{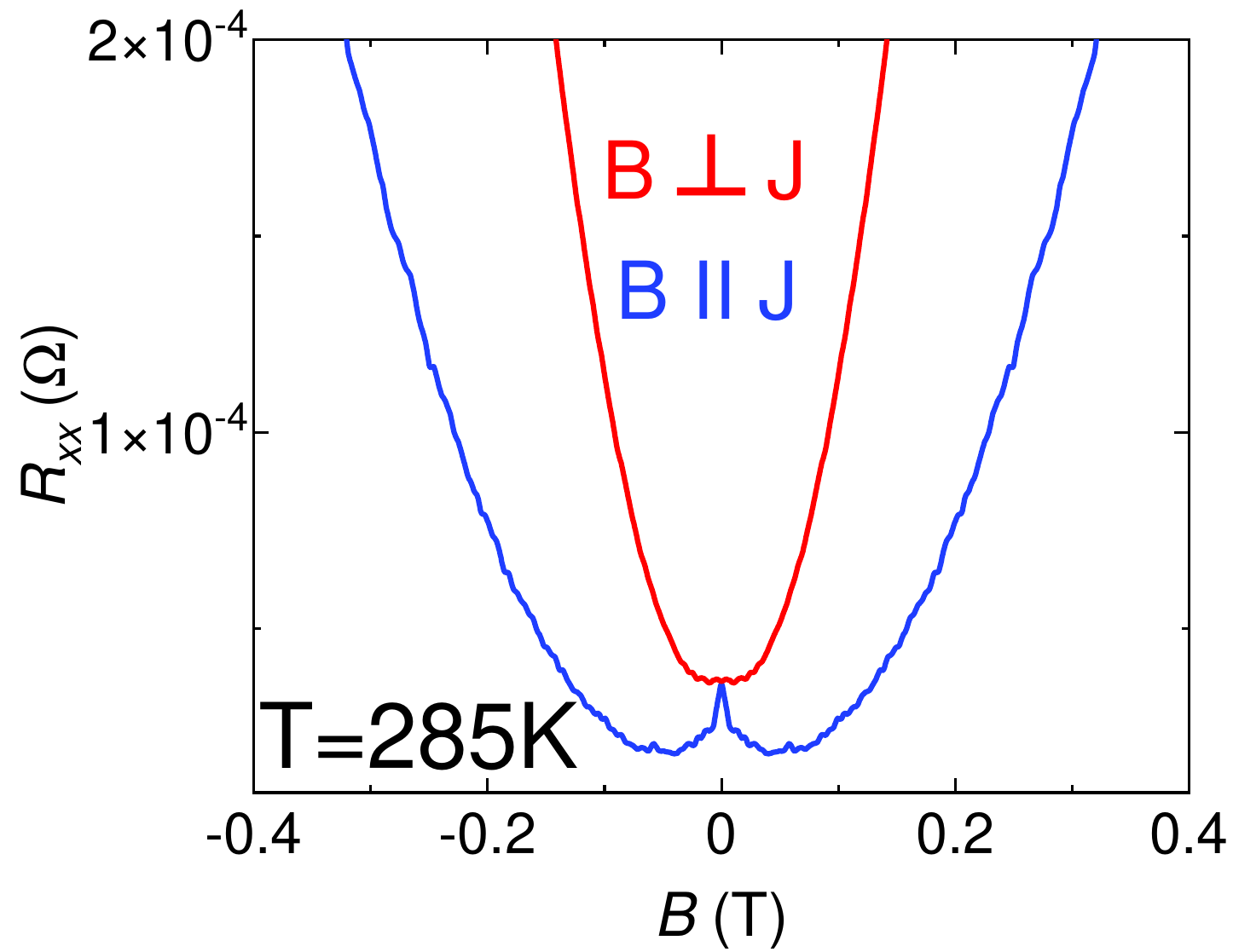}%
	\caption{\label{fig:CA} Magnetic field dependent longitudinal resistance $R_{xx}$ for Pb$_{0.58}$Sn$_{0.4}$Cr$_{0.02}$Te sample. Two measurement configurations are presented – with parallel and perpendicular magnetic and electric fields, marked in blue and red, respectively.  }
\end{figure}

Figure \ref{fig:CA} presents a comparison of the resistance $R_{xx}$ as a function of magnetic field $B$ in two measurement configurations – with electric field applied perpendicular and parallel to the magnetic field. The presence of chiral anomaly in the parallel configuration supports the interpretation that our system is a 3D WSM. Additionally, it is not related to the weak antilocalization – weak localization transition, since if this were the origin of our results,  it would be present as well in the parallel and perpendicular configurations.

\section{Absence of current jetting }
\begin{figure} [htbp]
	\includegraphics[width=1\textwidth]{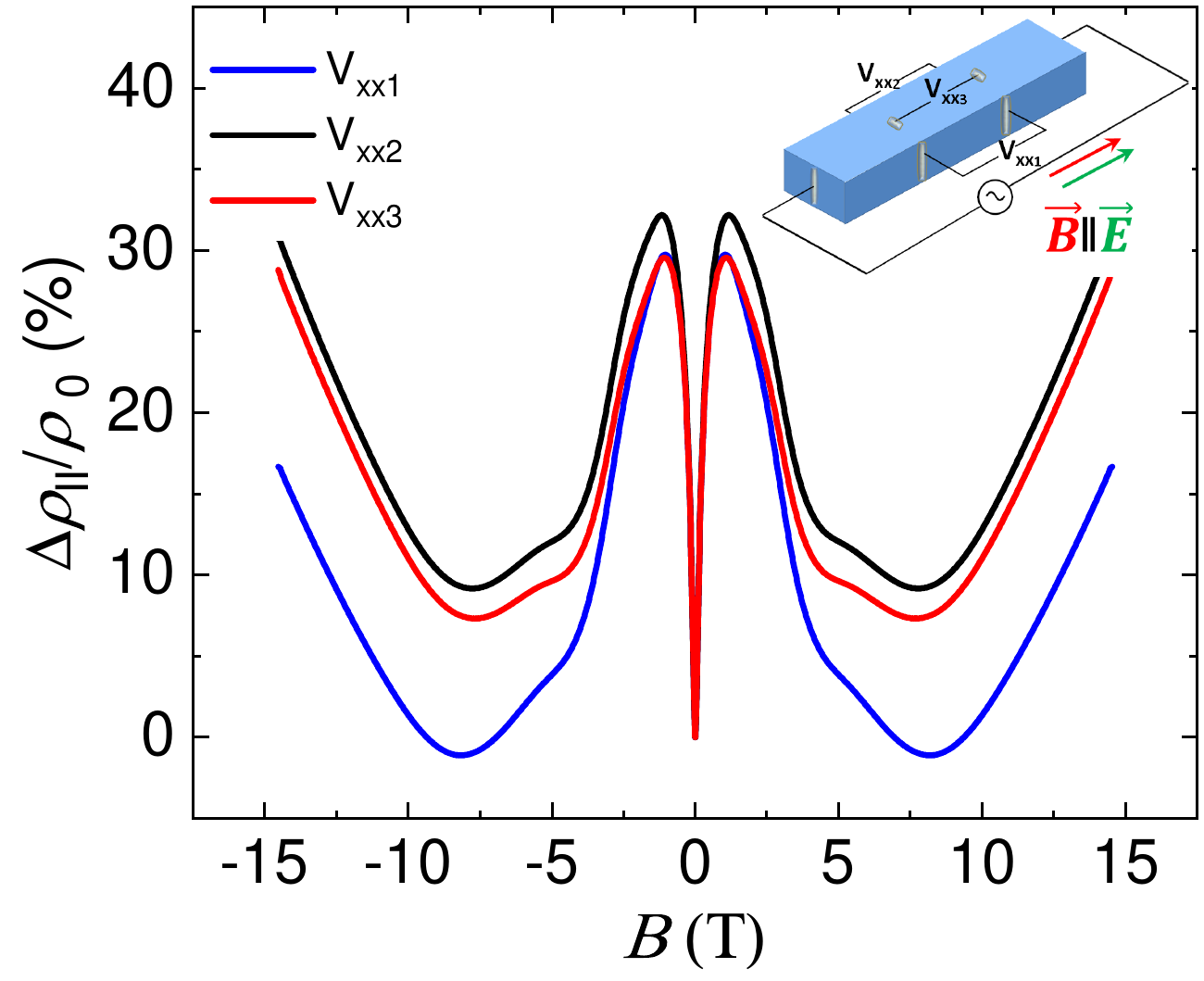}%
	\caption{\label{fig:curjet} Longitudinal resistivity versus magnetic field dependence for sample Pb$_{0.58}$Sn$_{0.4}$Cr$_{0.02}$Te. The signal measured on three pairs of contacts simultaneously exhibited similar response with the chiral anomaly on each pair, revealing the absence of current jetting.  }
\end{figure}

We also revealed the absence of the, so called, current jetting in our samples, which could adversely affect the results, since this is an effect related to the non-uniform distribution of the electric field (Fig. \ref{fig:curjet}).

\newpage
\section{Comparison of DC and AC measurement techniques }

\begin{figure} [htbp]
	\includegraphics[width=1\textwidth]{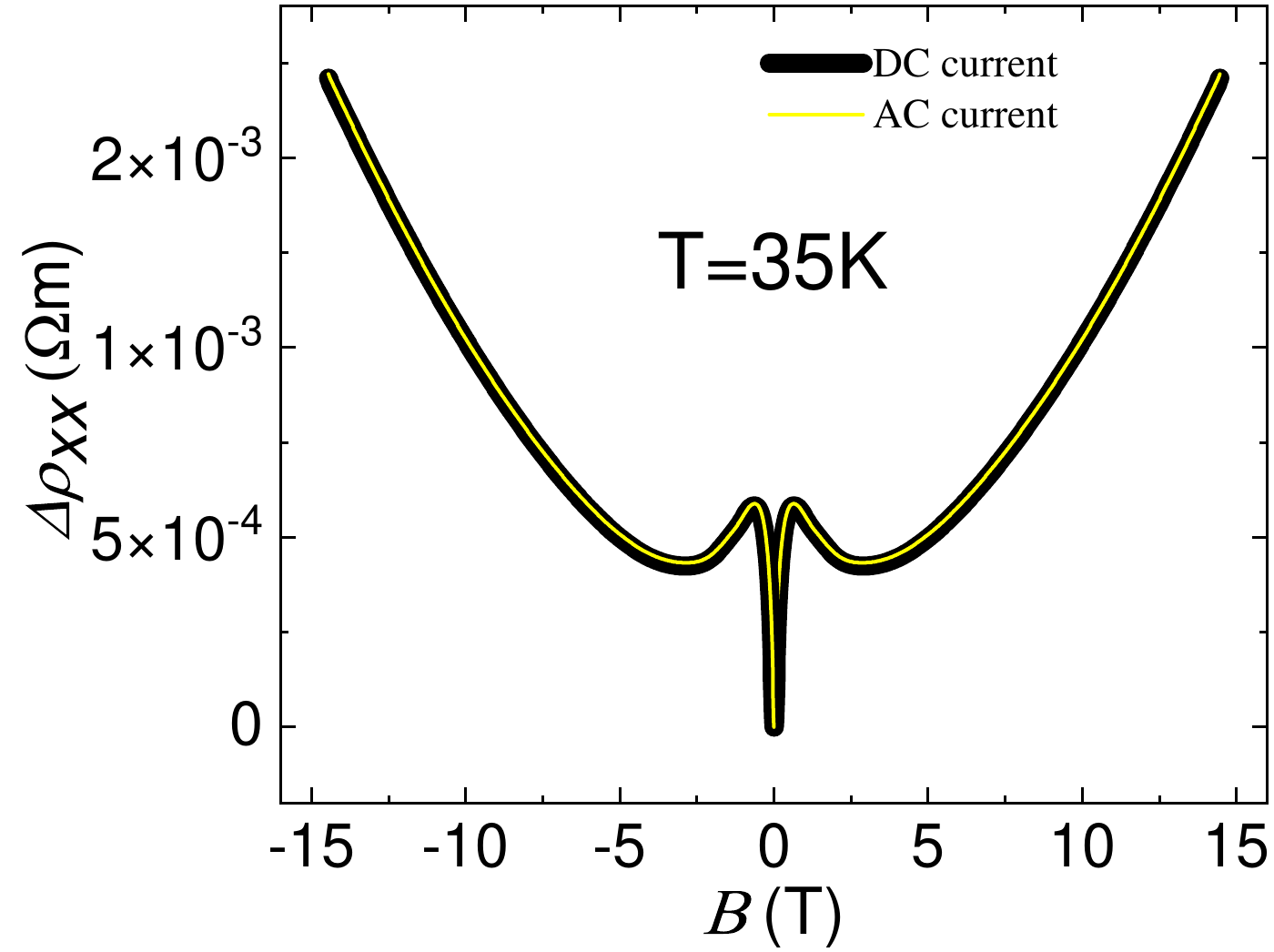}%
	\caption{\label{fig:ACDC} Comparison of the two measurement techniques – DC (black) and AC (yellow) current. The results obtained using both techniques are identical, confirming that the chiral anomaly underlies the observed effects. \cblu{It also excludes thermoelectric effects, such as Peltier or Seebeck effect.} }
\end{figure}
\newpage
\section{The Aharonov-Bohm and Altshuler-Aronov-Spivak oscillations }
\begin{figure} [htbp]
	\includegraphics[width=1.1\textwidth]{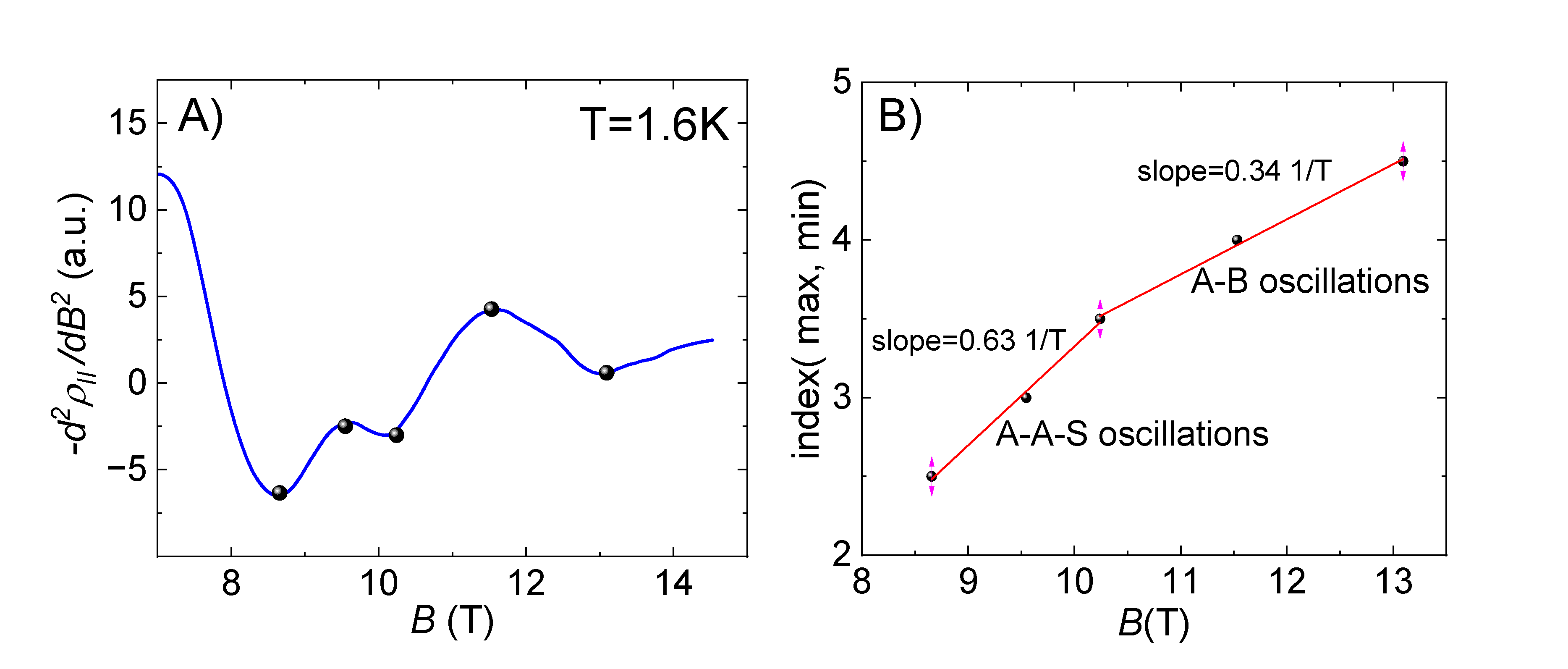}%
	\caption{\label{fig:ABAS} Oscillations periodic in the direct magnetic field: Aharonov-Bohm and Altshuler-Aronov-Spivak. (A) Second derivative of the resistivity over the magnetic field measured in the perpendicular direction. (B) Index plot as a function of magnetic field showing both – A-A-S and A-B oscillations, differing in the frequency. }
\end{figure}

In addition to the SdH oscillations (periodic in $1/B$), we noted the presence of another type of oscillations - periodic in the magnetic field. These results are plotted in Fig. \ref{fig:ABAS}.
The oscillations periodic in the magnetic field arise  due to the interference of the amplitudes of the electron wave function as the electron encircles the closed loop surroounding the magnetic flux.
In the case of the Aharonov-Bohm (A-B) oscillations, two electron wave functions encircle a closed magnetic flux once. In the case of the Altshuler-Aronov-Spivak (A-A-S), two electron wave functions encircle the magnetic flux twice \cite{bergmann_weak_1984,figielski_mesoscopic_2000}. This gives the twofold  change in the phases of the amplitudes of the A-B and A-A-S oscillations. As a result, the flux period $\theta$ in the magnetic field equals $h/e$ or $h/2e$, respectively.  

There is also another alternative, considering oscillations periodic in the magnetic field, attributed to pseudospin Landau levels \cblu{in the regime of} ultra quantum limit\cite{wang_quantum_2020,ezawa_pseudospin-_2016}.
However, we exclude this scenario and adopt the simpler A-B and A-A-S interpretation. Our conclusions are supported by the existence of the insulating nanoprecipitates based on Cr-Te compounds in our samples. The  surface area calculated basing on the frequencies of the A-B and A-A-S oscillations is comparable with the dimensions of these inclusions and yields 100 nm${^2}$.

\end{document}